\documentclass[aps,pre,preprint,superscriptaddress]{revtex4-1}
\usepackage{graphicx}
\usepackage{amsmath}
\usepackage{amssymb}
\usepackage{mathrsfs}
\usepackage{eucal}
\usepackage{amsfonts}
\usepackage{color}
\usepackage{bm}

\newcommand{\etal}{{\it{et al.~}}}

\renewcommand{\d}{{\rm{d}}}

\begin{document}

\title{Cosmology in one dimension: Symmetry role in dynamics, mass oriented approaches to fractal analysis}
\author{Bruce N. Miller}
\email{b.miller@tcu.edu}
\affiliation{Department of Physics and Astronomy,Texas Christian University, Fort Worth, Texas 76129}
\author{Jean-Louis Rouet}
\affiliation{Univ d'Orl\'eans, ISTO, UMR 7327, 45071, Orl\'eans, France ;  CNRS/INSU, ISTO, UMR 7327, 45071 Orl\'eans, France ;  BRGM,  ISTO, UMR 7327, BP 36009, 45060 Orl\'eans, France}
\author{Yui Shiozawa}
\affiliation{Department of Physics and Astronomy,Texas Christian University, Fort Worth, Texas 76129}

\date{\today}

\begin{abstract}
The distribution of visible matter in the universe, such as galaxies and galaxy clusters, has its origin in the week fluctuations of density that existed at the epoch of recombination. The hierarchical distribution of the universe, with its galaxies, clusters and super-clusters of galaxies indicates the absence of a natural length scale. In the Newtonian formulation, numerical simulations of a one-dimensional system permit us to precisely follow the evolution of an ensemble of particles starting with an initial perturbation in the Hubble flow. The limitation of the investigation to one dimension removes the necessity to make approximations in calculating the gravitational field and, on the whole, the system dynamics. It is then possible to accurately follow the trajectories of particles for a long time. The simulations show the emergence of a self-similar hierarchical structure in both the phase space and the configuration space and invites the implementation of a multifractal analysis. Here, after showing that symmetry considerations leads to the construction of a family of equations of motion of the one-dimensional gravitational system, we apply four different methods for computing generalized dimensions $D_q$ of the distribution of particles in configuration space. We first employ the conventional box counting and correlation integral methods based on partitions of equal size and then the less familiar nearest-neighbor and k-neighbor methods based on partitions of equal mass. We show that the latter are superior for computing generalized dimensions for indices $q<-1$ which characterize regions of low density.
\end{abstract}


\maketitle

\section{Introduction}

The analysis of large scale surveys shows that the grouping of visible matter presents a hierarchical structure on very large scales \cite{peebles2}. Stars are grouped into galaxies, galaxies into clusters, clusters into super-clusters, and so on. The shape of the ultimate element of this hierarchy has yet to be determined. Recently, a new grouping technic using the peculiar velocities of galaxies led to the discovery of the Laniakea supercluser, thereby extending the structure horizon \cite{Tully:2014aa}. This hierarchical regrouping, for which the enlargement of a structure reveals the existence of a smaller one, suggests a fractal arrangement of matter\cite{Martinez,Fang}. A true mathematical fractal object exhibits structures on all scales \cite{Man}; however, in practice, for the case of a physical object, there are both upper and lower bounds to the scaling structure \cite{feder88}. A quantitative measure of the range of scales that accommodate a hierarchical (fractal) distribution in the universe is provided by the correlation function of the distribution of galaxies obtained from recent surveys \cite{Martinez}.

Numerical simulation has made a significant contribution to the study of structure formation. It allows us to follow the dynamical evolution of these structures that change too slowly to be observed \cite{Virgo}. A number of specific, three dimensional, hydrodynamic and N-body codes are employed. However they require considerable resources and are limited in their ability to reveal fractal structure by the finite resolution that can be realized \cite{Alimi:2012be}.   Alternatively, by limiting simulations in $\mu$ space to one space dimension and an $N$-body description, we can increase the number of particles per dimension and treat the dynamics exactly, thereby retaining all the information concerning the particle trajectories. Of course, the finite size of the memory that represents numbers on computers doesn't allow the strict reversibility of the trajectories.

A purely one dimensional model dates from the year 1990 \cite{Rouet1}. The spherical version (hereinafter the Q model) was introduced by Fanelli and Aurell \cite{Fanelli}. An infinite version of the system was studied by Gabrielli \etal including the temporal evolution of the power spectra \cite{Joyce_1d}. The multifractal properties of a periodic version have been studied by Miller and Rouet  \cite{millerrouet,ewald}. Joyce \etal have studied the virialization of the clusters with regard to three dimensional observations. Virialization of gravitational systems have been given in numerous studies, for example \cite{HohlFeix,Yawn}. The fractal properties of these structures have notably been placed in evidence by Tsuchiya et al. \cite{tsuchiya2}.

In the earliest studies of the one-dimensional system \cite{Rouet1}, the initial condition was chosen as a ``water bag'' in which the particles were equally spaced in position, but their velocities were chosen independently and at random from a uniform distribution symmetric about the origin. After the system evolved, a fractal analysis was performed using the box-counting dimension. Depending on the Jeans' length associated with the initial condition, hierarchical scaling was revealed for the mass distribution in both phase space and configuration space.  Our more recent collaborative efforts have followed three principle avenues: the investigation of the possible existence of multifractal structures, the influence of scale free initial conditions closer to those following inflation revealed by WMAP, and the influence of changes in the parameters of the one-dimensional model \cite{MRGexp,millerrouet}. In so doing we also showed how to rigorously formulate the evolution of a one-dimensional self-gravitating system obeying periodic boundary conditions \cite{ewald}.

In the present work, after considering how symmetry dictates the construction of the equation of motion of the one-dimensional gravitational system, we  focus on a comparison of different methods for carrying out the fractal analysis of the resulting distribution in configuration space. A development of our recent work is that the standard methods for computing generalized, or Renyi, dimensions $D_q$ exhibit problems for the low density regions characterized by negative values of $q$ \cite{millerrouet}.  Here we first present a derivation of the class of one-dimensional models that can be constructed based on symmetry arguments. We then select the model originally introduced by Rouet and Feix which is the most self-consistent of the class of one-dimensional models. After following the evolution from small initial perturbations in the Hubble flow to a highly clustered state, we compare mass oriented methods of fractal analysis with the results of more standard approaches based on partitions of equal size.  Following this introduction, in section \ref{model}, the family of models and algorithms for following the dynamics of the particles which compose the system are presented. A typical simulation is presented in section \ref{simulation} along with the results of the multifractal analysis. These results are discussed in the last section of the article.

\section{Models and algorithm}\label{model}

In the following we consider a segment of the universe with extension sufficiently small for the Newtonian approximation to remain valid. The system will be described by a set of N particles interacting pair-wise according to Newton's law of gravity. We assume that the distribution of matter is highly symmetric. Specifically, in three dimensional space, three models are considered for which the distribution symmetry is either planar, cylindrical or spherical. Let $x$ refer to the coordinate of a particle ($x$ corresponds to the radius of a cylinder in the cylindrical case for example). Its equation of motion is written

\begin{equation}
\frac{\d^2 x}{\d t^2}=E(x,t)
\label{eq1}
\end{equation}
where $E(x,t)$ is the gravitational field. To accommodate the expansion two new variables of space and time are introduced \cite{besnard:1123} where C(t) represents the cosmological scale factor \cite{peebles2}

\begin{eqnarray}
x &=& C(t)\, \hat{x} \nonumber \\
\d t &=& A^2(t)\, \d\hat{t}
\label{eq2}
\end{eqnarray}
where
\begin{eqnarray}
C(t) &=& (\gamma \omega_{J_0}t)^{\alpha}  \nonumber \\
A^2(t) &=& (\gamma \omega_{J_0}t)^{\beta}
\label{eq3}
\end{eqnarray}
$\omega_{J_0}$ is defined as the Jeans frequency,  $\omega_{J_0}^2=4\pi G \rho_0$ where $\rho_0$ is the mass density
taken at the initial time  $t=t_0$ at which the size of the  physical and rescaled universe correspond. So $C(t_0)=A(t_0)=1$ and thus
\begin{equation}
t_0=\frac{1}{\gamma \omega_{J_0}}
\label{eq3a}
\end{equation}

Assuming that the field in the new variables follows the usual Poisson law, the equation of motion in the rescaled variables is written:

\begin{equation}
\frac{\d^2 \hat{x}}{\d \hat{t}^2}+2A^2\left(\frac{\dot C}{C}-\frac{\dot A}{A}\right)\frac{\d \hat{x}}{\d \hat{t}}+ A^4\frac{\ddot C}{C}\hat{x}=\frac{A^4}{C^3}\hat{E}(\hat{x},\hat{t})
\label{eq1a}
\end{equation}
where  $\dot C$ notes the derivative of $C$ with respect to time $t$ and $\ddot C$ the second derivative. In taking $\alpha=2/3$ and $\beta=1$, equation (\ref{eq1a}) no longer contains any time dependent coefficients. It becomes

\begin{equation}
\frac{\d^2 \hat{x}}{\d \hat{t}^2}+ \frac{1}{3} \gamma \omega_{J_0}\frac{\d \hat{x}}{\d \hat{t}}-\frac{2}{9}(\gamma \omega_{J_0})^2\hat{x}=\hat{E}(\hat{x},\hat{t})
\label{eq4}
\end{equation}

With the choice $\alpha=2/3$, $C(t)$ represents the Hubble expansion for a matter dominated universe that is appropriate for the epoch just following recombination and this is what we assume here \cite{peebles2}. Then the new variables represent a comoving frame and only the residual movement will be calculated. Note that the new unit of time $\d \hat{t}$ is now constant:

\begin{eqnarray}
\omega_J(t)\d t &=& \sqrt{4\pi G \rho(t)} \d t \\
                &=&  \omega_{J_0} \d \hat{t}
\label{eq5}
\end{eqnarray}

The transformation (\ref{eq2}) reveals two new terms in the equation (\ref{eq1a}), a friction and a force proportional to $\hat{x}$. In the following, it will be shown that the coefficient $\gamma$ depends on the type of symmetry and that, with one exception, the latter term of the left hand side of equation (\ref{eq1a}) must be modified so as to neutralize a uniform system that exactly follows the expansion. Under these conditions, the rescaled system will be static. It is a planar perturbation of this static universe that we follow hereafter.

\subsection{Symmetries and dimensions}

From the imposed symmetry, i.e., given that the field depends only on  $\hat x$, in $d$ dimensions the divergence operator is written:
\begin{equation}
\frac{1}{\hat{x}^{d-1}}\frac{\d \hat{x}^{d-1}\hat{E}}{\d \hat{x}}
\label{eq4.0}
\end{equation}

where $d = 1$ for planar symmetry, $d = 2$ for cylindrical symmetry, $d = 3$ for spherical symmetry, etc.. From the Poisson equation, in the comoving frame the scaled field then becomes

\begin{equation}
\hat{E}=-\frac{4}{d}\pi G\hat{\rho}\hat{x}.
\label{eq4.1}
\end{equation}
for the mean Hubble flow as $\hat{\rho}$ is constant. In that case the first two terms of equation (\ref{eq4}) vanish. By requiring that the force is proportional to $\hat{x}$, since the transformation compensates for that of gravity, we have

\begin{equation}
-\frac{4}{d}\pi G\hat{\rho}\hat{x}=-\frac{2}{9}(\gamma \omega_{J_0})^2\hat{x}
\label{eq4.2}
\end{equation}

and therefore

\begin{equation}
\hat{\rho}=\rho_0\quad \mbox{and}\quad \gamma=\frac{3}{\sqrt{2d}}.
\label{eq4.3}
\end{equation}
Equation (\ref{eq4}) can now be written in arbitrary dimension $d$ as

\begin{equation}
\frac{\d^2 \hat{x}}{\d \hat{t}^2}+ \frac{1}{\sqrt{2d}} \omega_{J_0}\frac{\d \hat{x}}{\d \hat{t}}-\frac{1}{d}\omega_{J_0}^{\,2}\hat{x}=\hat{E}(\hat{x},\hat{t})
\label{eq4.4}
\end{equation}

Considering our earlier assumption that the static state (mean flow) is excited by a planar perturbation of amplitude $y$ we have

\begin{equation}
\frac{\d^2 \hat{y}}{\d \hat{t}^2}+ \frac{1}{\sqrt{2d}} \omega_{J_0}\frac{\d \hat{y}}{\d \hat{t}}-\omega_{J_0}^{\,2}\hat{y}=\hat{E}(\hat{y},\hat{t})
\label{eq4.5}
\end{equation}
The field is now that of a planar system so, in order to satisfy global neutrality (background + field), the coefficient $1/d$ has been removed from the background term. When $d = 1$, this seemingly arbitrary change is not necessary because both the disturbance and the system are planar. Following the notation of Fanelli \cite{Fanelli} this  model ($d=1$) will be called the RF model and the spherical model ($d=3$ ) the Q model (or Quintic). Note that, for a spherically symmetric system of dimension $d$, the relation (\ref{eq4.0}) remains valid. If $d$ approaches infinity, the coefficient of friction approaches 0. While this requires imagining a universe of more than 3 space dimensions, this observation gives a physical meaning to the model without friction, or H-model, which has also been studied \cite{millerrouet} elsewhere.

The system is composed of $N$ particles which are infinite parallel planes themselves, with equal and constant surface mass density  $\mu$ (with the effect of expansion taken into account by the scaling). They experience friction and are bathed in a fixed, neutralizing, homogeneous background. In this sense, this system is the inverse of a plasma, since here the particles are attracted to each other and repulsed by the background medium. In cosmology it is customary to consider a segment of the universe that is small compared to the Hubble distance, but large enough to contain many galaxies and, optimistically, the largest observed structures. Since the size is much less than the Hubble distance, relativistic effects can be ignored and Newtonian mechanics suffices to follow the evolution \cite{Newtap}.  In  order to minimize edge effects and more closely represent a segment of an infinite system, periodic boundary conditions (PBC) are assumed to apply \cite{Ber}. Analogously, for our one dimensional models, we consider a slice of the universe that also obeys periodic boundary conditions. Then the gravitational field due to a single plane must also include the contribution from the sum of all of its infinite replicas, commonly referred to in the literature as an Ewald sum \cite{ewald}.

Since the field due to an isolated sheet of mass is constant in space, obtaining the correct field for PBC is not a trivial problem. Elsewhere we have shown that each mass sheet (or particle) carries with it its own neutralizing background \cite{ewald}. This turns out to be the same background contribution that appears in the transformation to the comoving frame for the RF model without further adjustment. In this sense,  PBC are the only correct boundary conditions that are compatible. The field experienced by a particle in the primitive cell satisfying PBC is given by
\begin{equation}
E(x)=4\pi mG\left[\frac{N}{L}(x-x_{c})+\frac{1}{2}(N_{R}(x)-N_{L}(x))\right]\label{eq:totfield}
\end{equation}
where $2L$ is the system size and $x_c$ is the system center of mass computed in the primitive cell \cite{ewald}. The role of $x_c$ is to insure that no interruption of the field is experienced by an interior particle when a different particle crosses a cell boundary and enters from the opposite side.

Operationally, one way this Ewald sum can be achieved is by polarizing the system boundaries: each side acts as a reservoir of initially neutral particles that can be loaded with particles of effectively ``positive'' and ``negative'' mass. As a particle reaches an edge, this edge becomes positively ``charged'' and  the other one  is negatively ``charged'', thus introducing a compensating dipolar gravitational field. This technique is usually used in the case of a one-dimensional plasma and insures that there is no discontinuity in the field when a particle crosses a cell boundary. As explained in \cite{ewald} it is thus possible to follow the particle trajectories exactly until they intersect. Then, rather than the usual pattern of dynamical evolution, which is to advance molecular particles along their trajectory according a fixed time step, here we adopt an event driven scheme where the particles are advanced to their next crossing time with their neighbor. As the solutions of the equations of motion are known, it is possible to  calculate the shortest crossing time of a particle with its neighbor and then restart the process. For the RF (Quintic) case, this means solving a cubic (fifth order) equation and determining the root that yields the smallest positive crossing time. Thus the RF model has the advantage that the crossing times can be found analytically. For diagnostics, the particles are temporarily advanced to the current time. This is straightforward as the analytic solutions of the equations of motion between crossing events are known exactly. Typically the trajectories of $N=2^{18}$ particles are followed in quadruple precision. This precision is necessary to follow a large number of particles for a sufficient time to  enable  large size structures to grow, while retaining a high precision for the distribution of structures of smaller size (cf. figure \ref{fig1}).

\section{Multifractal Analysis of Simulations}\label{simulation}
Following inflation, density fluctuations in the universe can be modeled as a Gaussian random field with a scale-free (power law) power spectrum \cite{peacock}. In the three dimensional universe, the exponent corresponding to a scale-free potential is unity \cite{peacock}. Initial conditions for 3D simulations of the expanding universe are guided by these considerations. For the simulation data reported here, we selected the RF model and, in order to achieve scale-free potential fluctuations in 1D, the power spectrum of the density fluctuations at the initial time is chosen to vary as $k^{3}$, where $k$ is the wave number. The construction of the initial distribution of the particles allowing us to sample this spectrum is given by Miller et al. \cite{millerrouet}. In the past, other initial conditions were adopted, particularly the water bag model for which the particle velocities are distributed randomly, following a uniform distribution between $\pm a$, while the positions are equally spaced. If the friction term is omitted, it is possible to write a dispersion equation which shows that the system is unstable for wave numbers $k$ such that $k>k_{c}=2\pi/\lambda_J$ where $\lambda_J$ is the Jeans length. This particular initial condition was  used to set the Jeans length  $2\pi a/\omega_J$. Nevertheless, for the initially cold system simulated here, the Jeans length tends to 0, and these initial conditions, with a power law spectrum of the density, do not favor any particular length scale. However, care must be taken in the choice of the spectral index \cite{peacock}. If we take $k^{-1}$ for example, we will give too much energy to large scales and we will just arrive at one large cluster, perhaps with a few small ones, while we require $k\leq 4$ to guarantee momentum conservation \cite{peacock}.

\begin{figure}[ht]+
\centerline{\includegraphics[width=\textwidth]{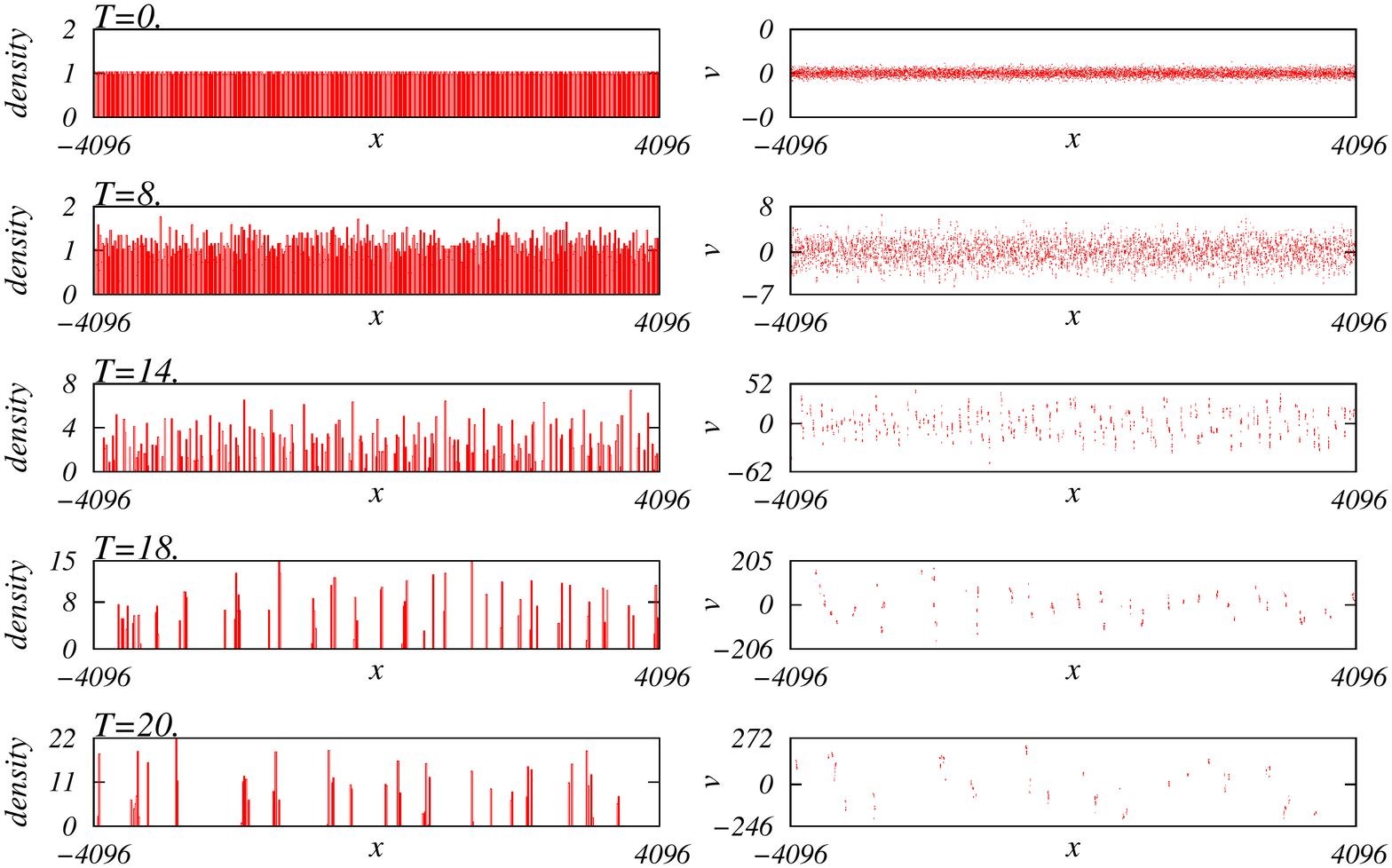}}
\caption{\label{fig1}Snapshots of the distribution of particles in configuration and phase space for a system of  $N=8191$ particles at the instants $T=0$, 8, 14, 18 and 20.}
\end{figure}

\begin{figure}[ht]
\centerline{\includegraphics[width=\textwidth]{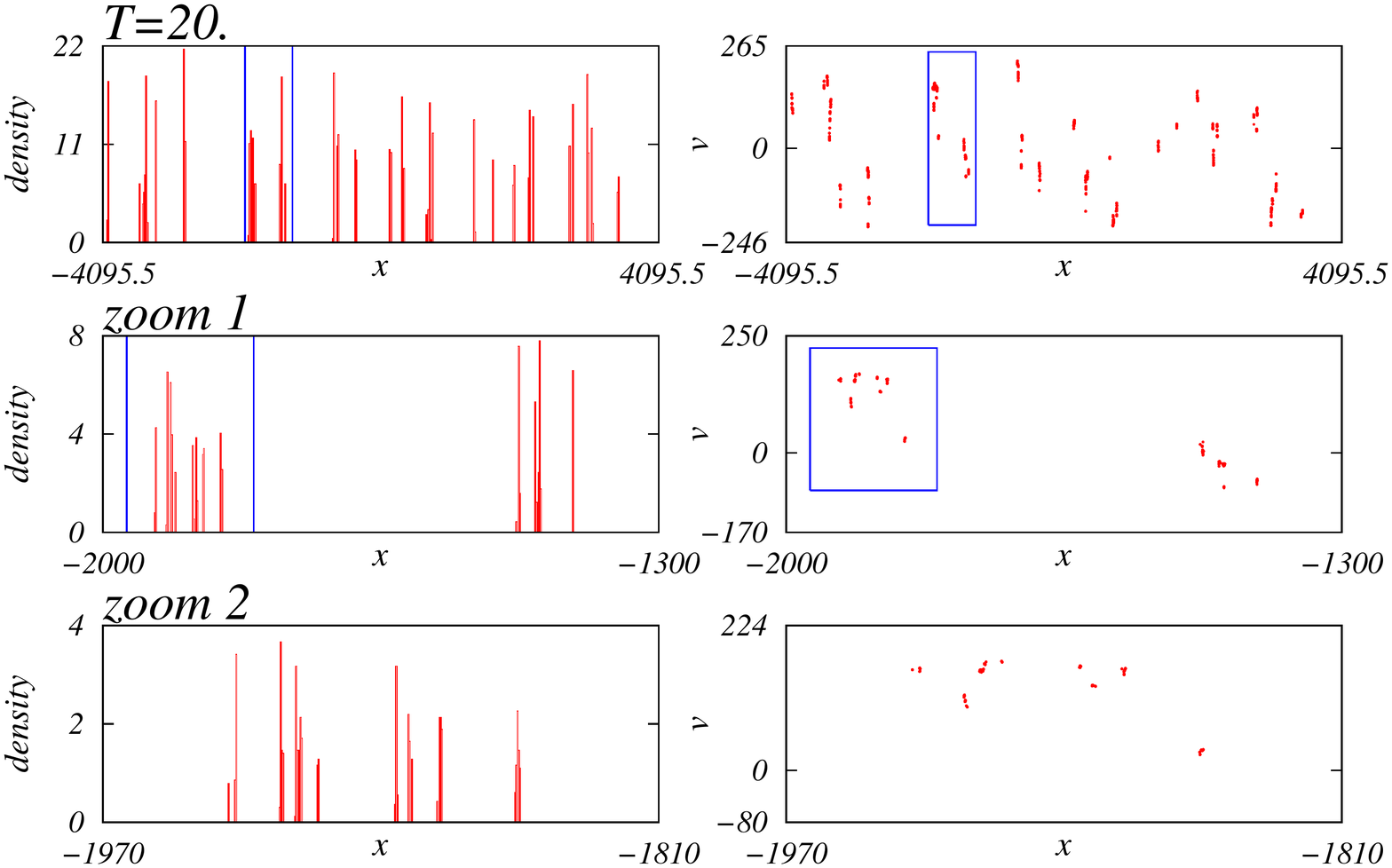}}
\caption{\label{fig2} Two successive zooms of a patch of phase space (upper figure) at $T=20$ . The enlarged areas are bounded by dashed lines.}
\end{figure}

Figure \ref{fig1} shows a sequence of snapshots of the density and distribution of $N\sim2^{13}$ particles in the phase space. The initial fluctuations cause the system to rapidly enter  a non-linear regime wherein structures of small size appear. These structures locally regroup to form larger structures that are themselves regrouped, and so on. Two successive enlargements of a larger structure formed at time  $T=20$ are shown in figure  \ref{fig2}. It can be seen both in the phase space, but also on the density peaks, that these large structures are formed of smaller structures, themselves resulting from an assembly of structures of less size.

This hierarchical structure suggests that, at least over a finite range of scales, the system is fractal. To test this hypothesis, a multifractal analysis on the distribution of particles in the configuration space is conducted using four methods. Fractals can be characterized by their dimension. For simple fractals, such as a Koch curve or the Cantor set, a single number is sufficient, whereas for systems that are less homogenous, a continuum of dimensions is required. Considering the large variation in density exhibited on the line and in the phase space in the results of our simulations, it would seem that the latter approach has to be pursued.

Renyi introduced the idea of generalized dimension $D_{q}$  by first partitioning the embedding space into $N_b$ equal size cells and determining the measure $\mu_i$ associated with each. For the case of our simulations we consider the local density $\mu_i=N_i/N$ within each box where $N_i$ is the number of particles in this box. Let $C(q,l)$ be the partition function identified with a particular decomposition:
\begin{equation}
C(q,l)=\sum_{i=1}^{N_b(l)}\mu_i^q(l)\quad \mbox{\rm and}\quad   \lim_{l \rightarrow 0} C(q,l) = l^{\tau_q}.
\label{req-1}
\end{equation}
Then the generalized or Renyi dimensions are defined by
\begin{equation}
\left\{
\begin{array}{ll}
\displaystyle D_q=\lim_{l\rightarrow 0}\frac{1}{q-1}\frac{\ln(C(q,l))}{\ln(l)} & \mbox{\rm for } q\ne 1 \\
\displaystyle D_q=\lim_{l\rightarrow 0}\frac{\sum_{i=1}^{N_{b}(l)}\mu_i\ln(\mu_i)}{\ln(l)} & \mbox{\rm for } q=1 \\
\end{array}\right.
\label{req-2}
\end{equation}
The exponent $ \tau_q$ is related to the Renyi dimension of order $q$ by $D_q=\frac{\tau_q}{q-1} $. Note that, from the definition, regions of high density contribute strongly for positive values of $q$, and low density regions more strongly support negative values.

In practice, for simulated or experimentally observed data, it is not possible to go to the limit that the definition demands. To compensate, two general methods for determining fractal dimensions have evolved. In one, the space in which the fractal set is embedded is partitioned into subsets of equal size, whereas, in the second, it is partitioned into sets of equal measure or mass. The first method follows more closely from the definition of Renyi Dimension. It is realized in both the box counting and correlation methods, which are the most popular for estimating dimensions of natural sets. The second approach is realized in the near-neighbor \cite{Badii85} and k-neighbor \cite{Water88} methods and has the advantage of only including occupied cells in the partition. Recently, to gain insight concerning their useful regimes, we have applied these methods to standard sets with well-characterized fractal properties \cite{yui}.  To paint as clear a picture as is possible regarding the fractal nature of the gravitational simulations, here we will pursue each approach and see what further insights they provide.

In the box counting (BC) method, since the limit in Eq.~(\ref{req-2}) cannot be performed in practice, rather plots of $\ln(C_q)$  vs $\ln(l)$  are studied to determine if linearity is present, suggesting power law behavior in the partition function that is required for a well defined limit. We operationally define the generalized dimensions using
\begin{equation}
\left\{
\begin{array}{ll}
\displaystyle D_{q,l}=\frac{1}{q-1}\frac{\ln(C(q,l))}{\ln(l)} & q\ne 1 \\
\displaystyle D_{q,l}=\frac{\sum_{i=1}^{N_{b}(l)}\mu_i\ln(\mu_i)}{\ln(l)} & q=1 \\
\end{array}\right.
\label{req-3}
\end{equation}
The exponent $ \tau_q$ is related to the Renyi dimension of order $q$ by $D_q=\frac{\tau_q}{q-1} $.
The representation of $\ln(C(q,l))/(q-1)$ as a function of $\ln (l)$ gives the value of $D_q$ in the zone for which the plot is linear. Note that if $l$ is too large, then the slope is equal to 1 which indicates that the system is homogeneous on this scale. Conversely if $l$ is too small, then the slope is zero, which is an effect of the discretization of the system.
Due to the simplicity of the method, it is widely used among researchers. However, it has been pointed out by many that this method and, more generally, methods that involve partitions of the same size such as the correlation method, do not work well for $q < 1$.  \cite{Riedi95,yui}
In the correlation integral (CI) method introduced by Grassberger and Procaccia \cite{Itamar83},  one fixes on a selected set of reference points and examines the distribution of points in neighborhoods of equal size surrounding each of them. This provides an alternative formulation for a partition function $I_{q}(l)$ for a partition of the system with cells of equal size. Let
\begin{equation}
I(l)=\frac{1}{N}\sum_{i=1}^{N}\mu_{i}^{q-1}(l)
\end{equation}
where, here,
\begin{equation}
\mu_{i}=\frac{1}{N}\sum_{j=1}^N\theta(l-|x_i-x_j|)
\end{equation}
where $\theta(x)$ is the unit step function and
\begin{equation}
D_{q}=\lim_{q \to 0}\frac{1}{q-1}\frac{\ln{I(l)}}{\ln(l)}
\end{equation}
as before. Then the generalized dimension  ${D_q}$ can be obtained in similar fashion to the BC method.

	To formulate approaches based on partitions of equal measure or mass, let $\delta_{j}(k,n)$ be the distance between a point $x_j$ chosen from our set and the  $k^{th}$ nearest neighbor to $x_j$. Next construct the sum of the moments $\delta_{j}(k,n)$ from a set of $n$ reference points chosen at random:

\begin{equation}
\Delta^{(\gamma)} (k,n) = \frac{1}{n} \sum_{j=1}^{n}\delta_j^\gamma(k,n). \label{eq: k-1}
\end{equation}
Van der Water and Schramm have shown that \cite{Water88}, in the limit of large $n$,
\begin{equation}
\left<\Delta^{(\gamma)}(k,n) \right>^{1/\gamma}  \cong n^{-1/D(\gamma)} \left[\alpha D(\gamma) \frac{\Gamma(k+\gamma/D(\gamma))}{\Gamma(k)} \right]^{1/\gamma}\cong n^{-1/D(\gamma)}K(\gamma,k)
\label{eq: k-3}
\end{equation}
where $\alpha$ is a constant independent of $\gamma$ and the Dimension Function, $D(\gamma)$, can be thought of as alternative generalized  dimension. It is related to the Renyi dimension through
\begin{equation}
D[\gamma = (1-q)D_q] = D_q.
\label{req-5}
\end{equation}
Note that the average of $\delta^{\gamma}_{j} $ from a single set is used in Eq. (\ref{eq: k-1}) whereas the derivation of Eq. (\ref{eq: k-3}) is based on the ensemble probability.

Using their result, two different approaches for determining $D(\gamma)$, and therefore $D_q$, can be extracted.  For a fixed value of $k$, the partition elements all consist of intervals with the same number of particles. One can then investigate the dependence of $\Delta^{(\gamma)}$ on the sample size $n$ and obtain $-1/D(\gamma)$ from  the slope of plots of $\log(\Delta^{(\gamma)}/K)$ vs $\log( n) $.  In doing so note that care must be taken to avoid the singularities of the Gamma function \cite{abramowitz70}.
For the special case where $k=1$ this is known as the nearest-neighbor (NN) method and was extensively investigated by Badii and Politti \cite{Badii85}, later by Broggi \cite{broggi1988evaluation} and, very recently, by us  \cite{yui}. However, it has also been applied with fixed values of $k$ as large as 300 \cite{broggi1988evaluation}.
	Alternatively, we can fix the number of sample points $n$ and investigate the scaling of $\Delta$ with $k$. In so doing we are considering a cumulative sequence of partitions based on elements of increasing mass or measure. This approach is known in the literature as the k-neighbor (KN) method. Because it incorporates a wide range of $k$ values, it results in a more global measure of dimension and is less susceptible to local fluctuations that influence the other methods\cite{yui}. For large $k$, a simple approximate relation can be obtained:  \cite{Water88}
\begin{equation}
\left[ \Delta^{(\gamma)} (k,n) \right] ^{1/\gamma} \cong n^{-1/D(\gamma)} k^{1/D(\gamma)}G(k,\gamma) \label{eq: k-2}
\end{equation}
where $G(k,\gamma)$ is a correction function and is close to unity for large $k$.
The Dimension Function $D(\gamma)$ can be estimated by setting $G(k,\gamma)=1$ in the first iteration. The dependence of the correction function $G(k, \gamma)$ on $k$ and $\gamma$ can be obtained from Eq. (\ref{eq: k-3}) with the value of $ D(\gamma)$  from the first iteration. The Dimension Function $D(\gamma)$ is then updated using this $G(k, \gamma)$. After a few iterations, the numerical results for $D(\gamma)$ typically converge to a single value for each $\gamma$.

\begin{figure}[ht]
\centerline{\includegraphics[width=\textwidth]{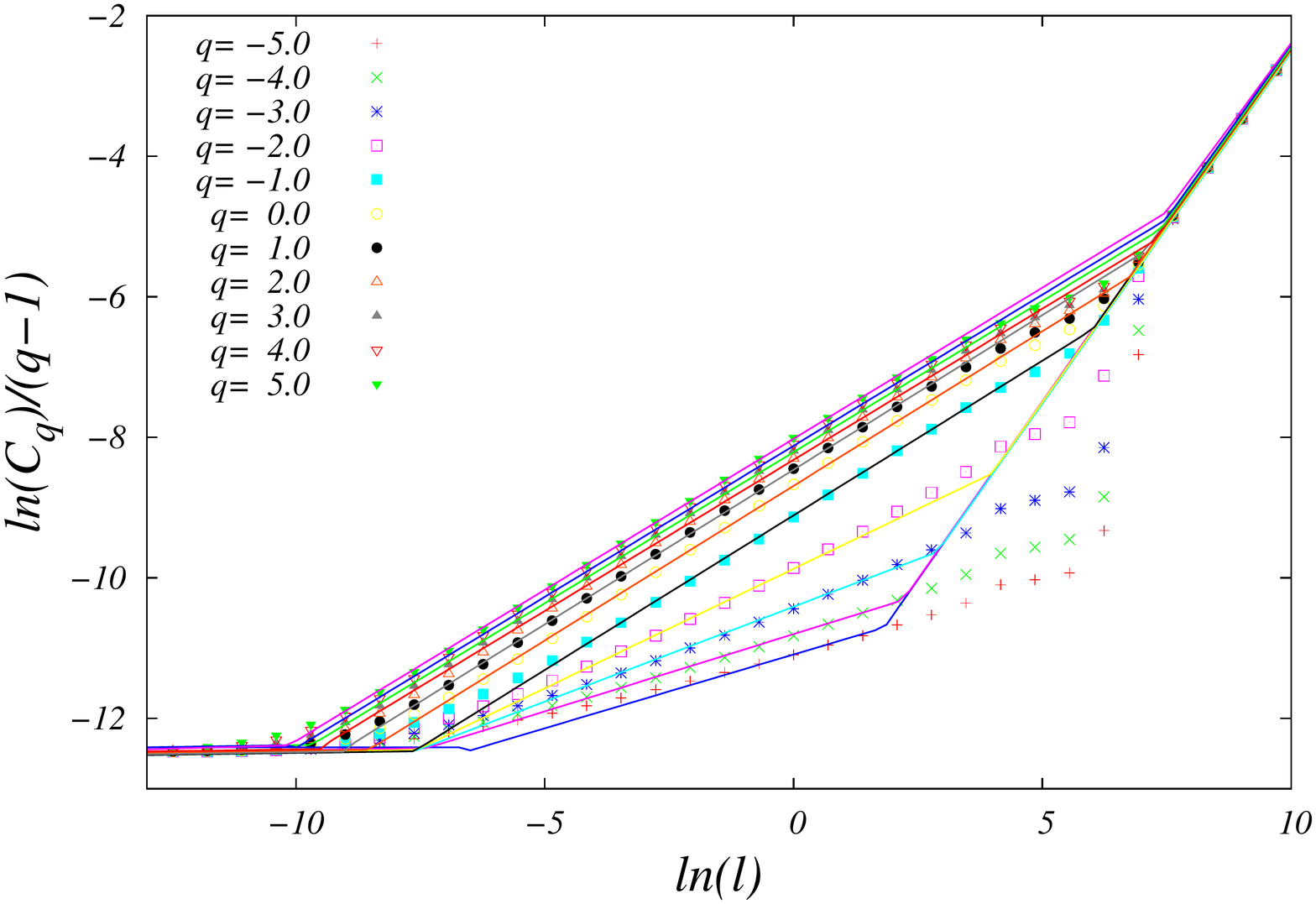}}
\caption{\label{fig3} BC method: $\ln(C_q)/(q-1)$ vs. $\ln(l)$ for several values of  $q$. The slope of the linear central zone gives $D_q$ (cf. figure \ref{fig4}).}
\end{figure}

\begin{figure}[ht]
\centerline{\includegraphics[width=\textwidth]{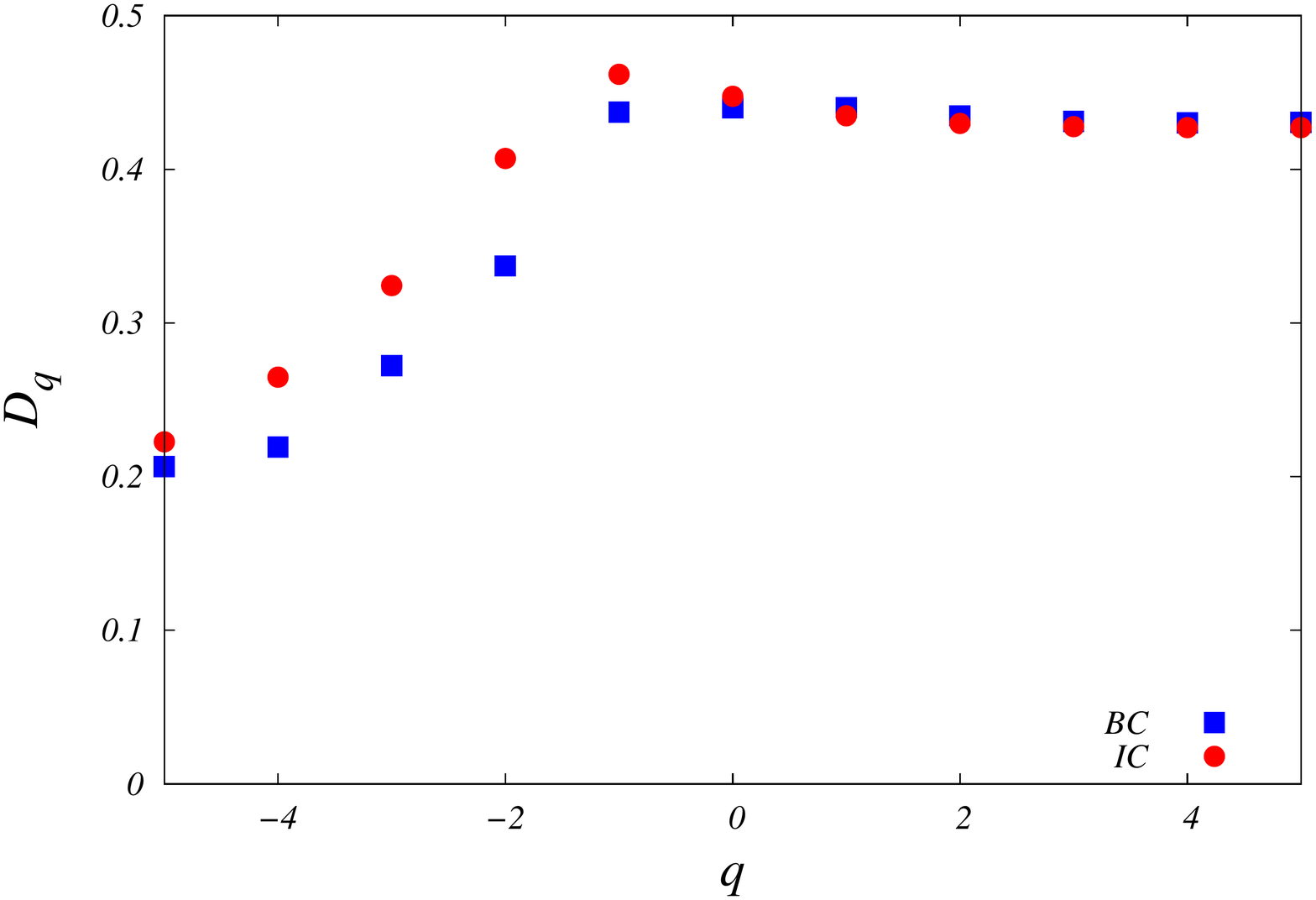}}
\caption{\label{fig4} Generalized dimension $D_q(q)$ obtained at $T=20$ by the BC (squares) and IC (circles) methods for a simulation of $N=2^{18}$ particles.}
\end{figure}

The set of analyses are carried out at time $T=20$ on a simulation with  $N=2^{18}$ particles. The particle distribution is similar to that of Figures \ref{fig1}  and \ref{fig2} for which the number of particles is $2^{13}$, except that the number aggregate is now 32 times greater, improving the statistics. Regarding the BC method, Figure \ref{fig3} shows a typical plot of $\ln(C_q)/(q-1)$ as a function of $\ln(l)$ for different values of $q$. You can notice the intervals where $D_q=0$  (zero slope) for  $l\rightarrow 0$ and where  $D_q=1$ for large values of $l$. Between these two zones, there is a third scaling interval which seems to expand with the value of $q$.  For $q = 0$, for example, it varies from about $10^{-3}$ to $10^{3}$, encompassing six decades. The values of the slope for these scaling regions are identified for several values of q. The function of $D_{q}$  vs $q$ thus obtained is shown in Figure \ref{fig4}. Similar results obtained by the CI method are also shown in Figure \ref{fig4}. The two curves are similar. For $q>-1$, they are decreasing, but increasing for $q<-1$. However, while theory  tells us that a general property of $D_q$ is that it is not strictly increasing, this is not the case here. This situation led us to employ mass oriented partitions to have better insight, especially for negative values of $q$.


\begin{figure}[htbp]
\centering
\includegraphics[width=\textwidth]{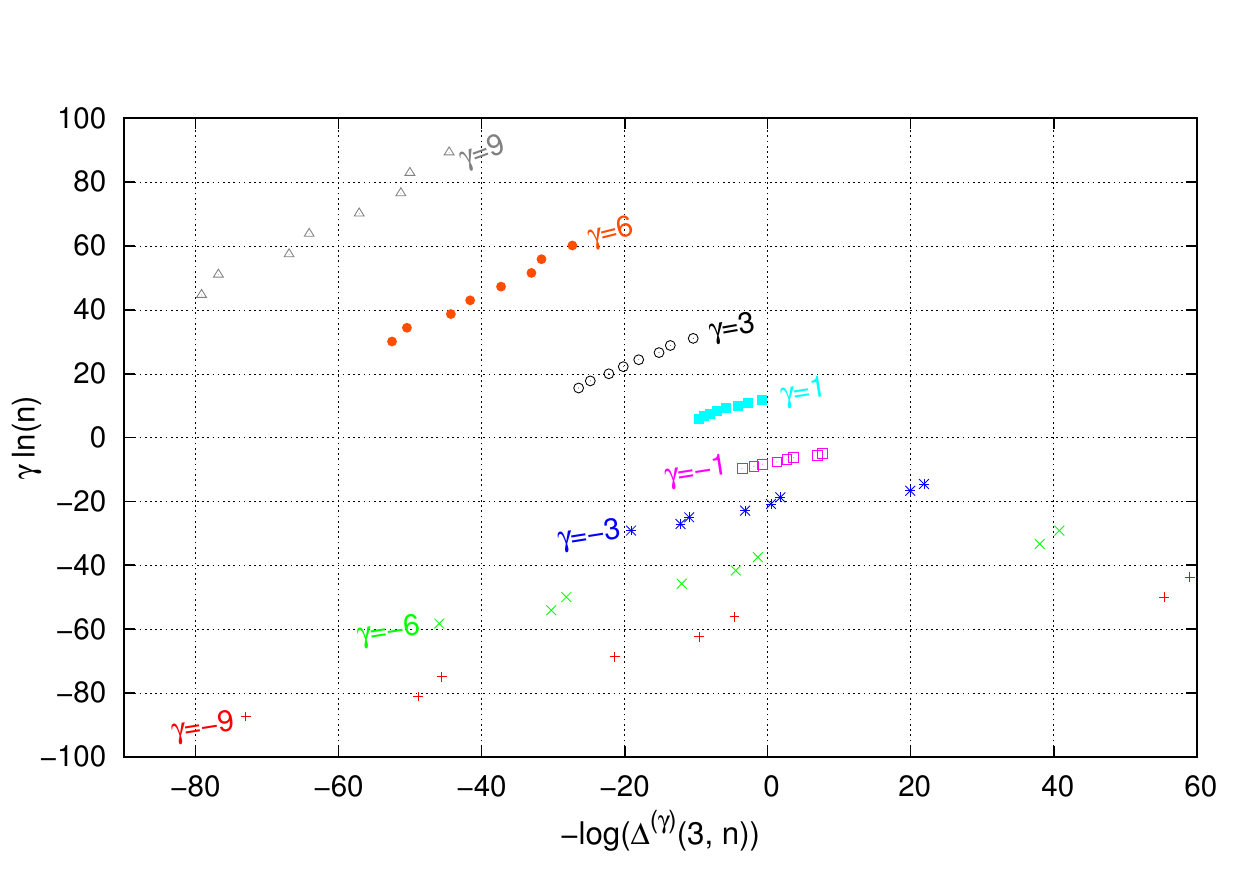}
\caption{{ This plot shows the scalings with the near-neighbor method. The slope of the best-fit line for each $\gamma$ is taken as the Dimension Function,  $D(\gamma)$.} }
\label{fig: Near-neighbor_scaling_T=20}
\end{figure}

As the nearest neighbor method is sensitive to the local statistical noise, we used $k=3$ instead of $k=1$ , which we refer to as the ``near-neighbor" method. As shown in Fig.~\ref{fig: Near-neighbor_scaling_T=20}, we have successful scaling for the positive range of $\gamma$. However, as $\gamma$ increases, the contributions from a few sample points start to dominate the sum in Eq.~\ref{eq: k-1}, and therefore the scaling becomes less stable. By increasing the fixed value of $k$, the location of the singularity is shifted further into the negative range of $\gamma$, thus increasing the size of the reliable range for this method.

\begin{figure}[htbp]
\centering
\includegraphics[width=\textwidth]{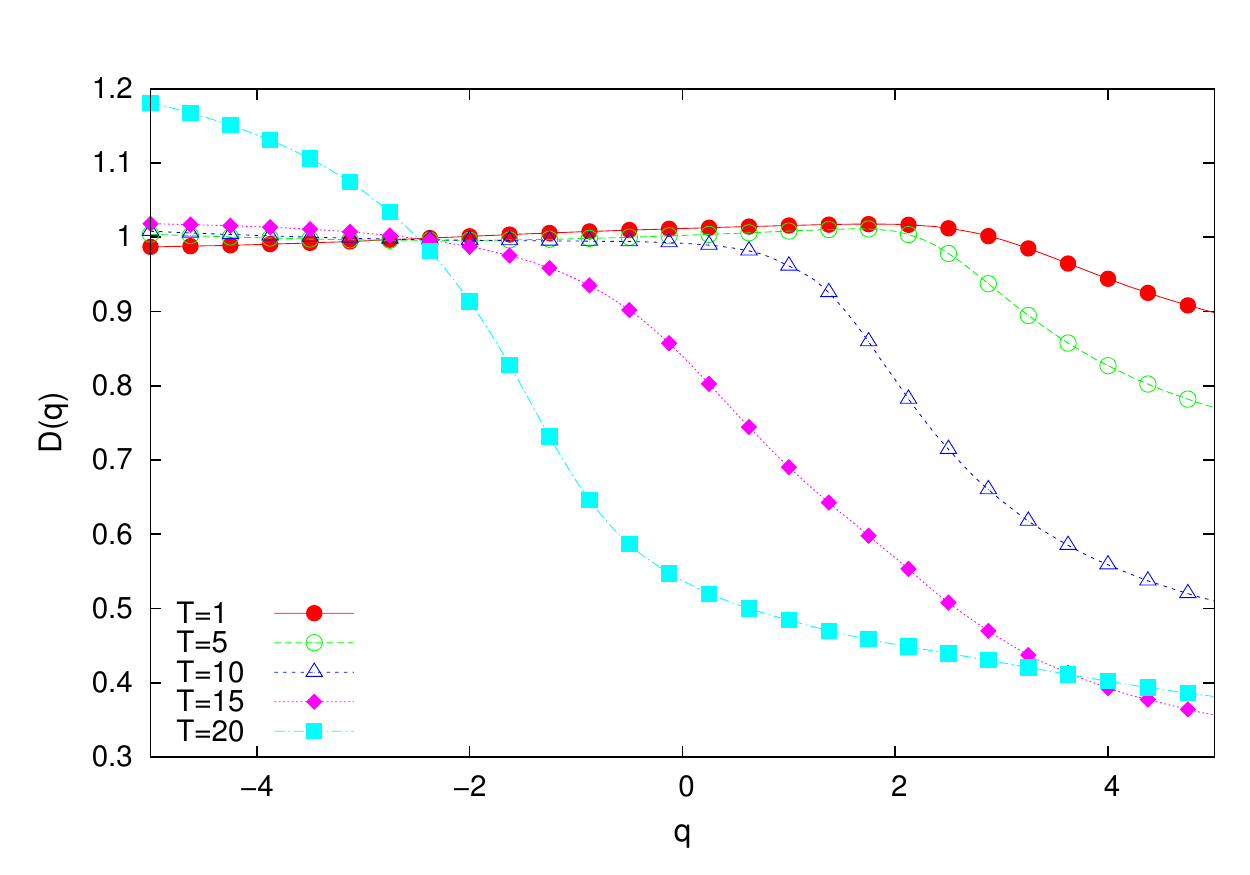}
\caption{{ This plots shows the generalized dimension using the third nearest neighbor method for different times $T$.}}
\label{fig: Dq_near-neighbor}
\end{figure}

The evolution of generalized dimension $D_q$ over time $T$ computed with the near-neighbor method is shown in Fig.~\ref{fig: Dq_near-neighbor}. At the beginning of the simulation, we obtain a constant $D_q$ with a value close to 1 as expected. For the range $q > 2$, the generalized dimension starts to diverge from the expected spectrum. Accordingly, we conclude that the near-neighbor method with $k=3$ is not reliable for $q>2$. This conforms with the analysis performed on sets with known fractal properties \cite{yui}. For the range where this method is reliable, we obtained a smooth, non-increasing, spectra for $D_q$. For the range $q > -2.5$, $D_q$ generally increases over time whereas for $q < -2.5$, the spectrum decrease over time. It is noteworthy that the generalized dimension is almost invariant over the entire duration of the simulation at $q=-2.5$ with $D_q = 1$. This observation may not be accidental and warrants further study.

\begin{figure}[htbp]
\centering
\includegraphics[width=\textwidth]{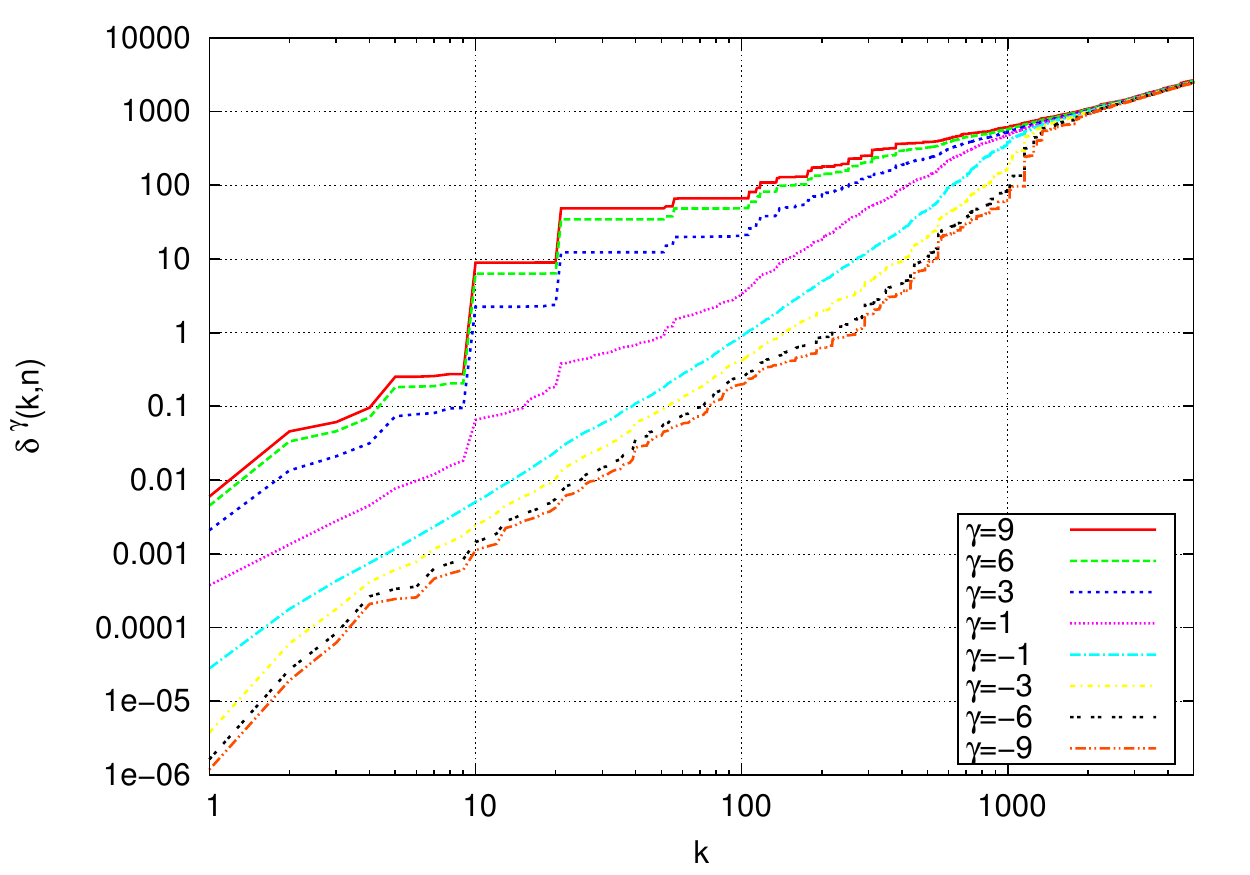}
\caption{{ This log-log plot shows the scalings with the $k$-neighbor method at $T=20$. }}
\label{fig: k-neighbor_scaling_T=20}
\end{figure}

\begin{figure}[htbp]
\centering
\includegraphics[width=\textwidth]{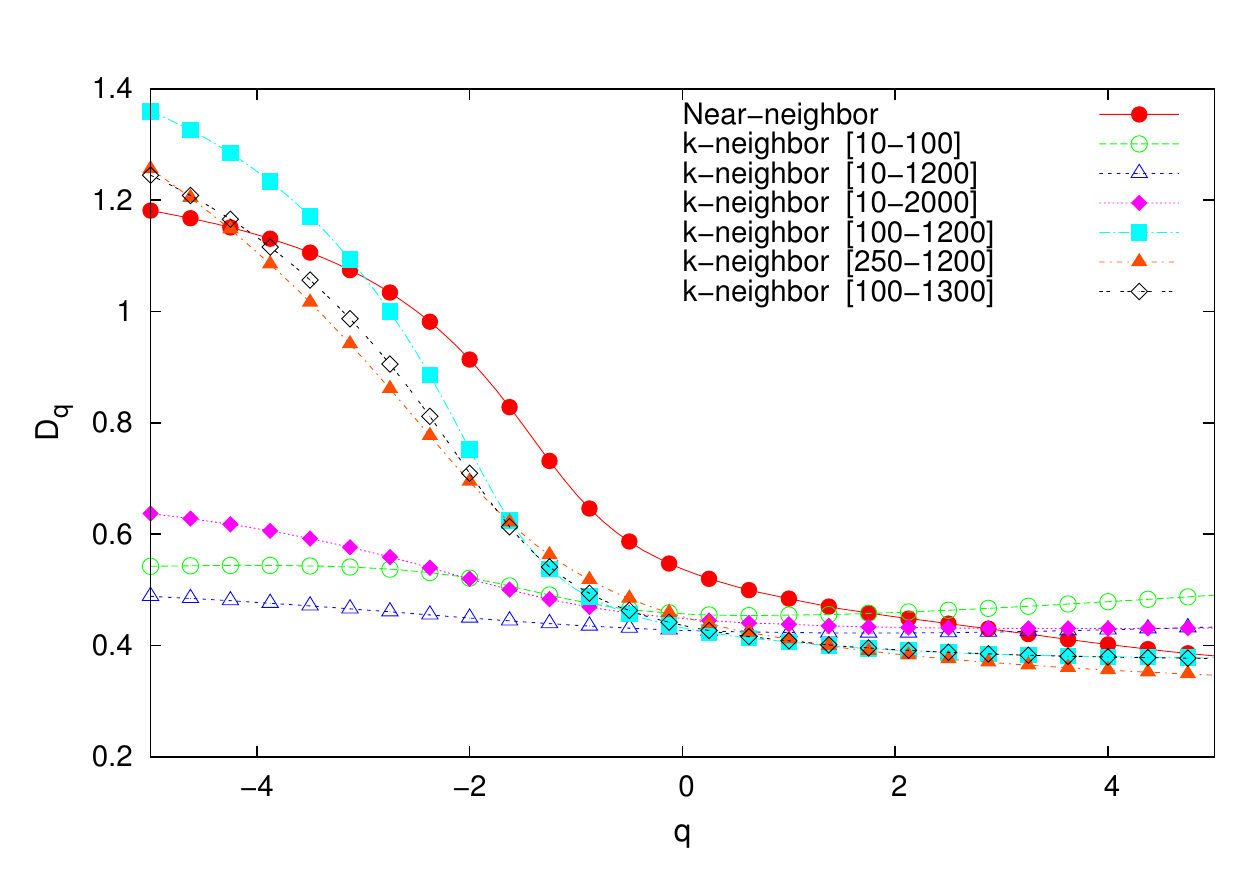}
\caption{{ This plot the sensitivity of the generalized dimension $D_q$ to the choice of the range from which the best-fit line is extrapolated in the $k$-neighbor method. The selected ranges are shown in the legend. The generalized dimension $D_q$ computed with the near-neighbor method is also included for comparison.}}
\label{fig: Dq_k-neighbor_T=20}
\end{figure}

To evaluate $D(\gamma)$ using the k-neighbor method, $\Delta^{\gamma}$, which has the dimensions of length, is plotted vs. the number of neighbors $k$ for several values of $\gamma$ on a log-log scale. From the slope, the values of $D_{q}$ for differing $q$ can be deduced following the relation (\ref{req-5}). In Fig.~\ref{fig: k-neighbor_scaling_T=20}, we show the plots obtained with the $k$-neighbor method. We divide the range of $k$ into four regions depending on the behavior of the weighted average of the $k$\textsuperscript{th} neighbor distance. In this particular plot, we see large gaps in the plots for positive $\gamma$ at, for example, $k=10$. The range $k<10$ is known to have singularities for the negative range of $\gamma$, and therefore we typically do not include this region in the subsequent computation.  The range $ 10<k<100$ exhibits typical fine structures which are the hallmark of a fractal. The gaps correspond to lacunarity in the simulated set. The generalized dimension $D_q$ extracted from this region is included in Fig.~\ref{fig: Dq_k-neighbor_T=20}. Although the sample points are limited, and therefore a careful analysis is required, the spectrum is more mono-fractal like with $D_q$ being significantly smaller than 1. The range $100<k<1200$ clearly has different slopes for different values of $\gamma$. The spectrum from this range is said to be multifractal as shown in Fig.~\ref{fig: Dq_k-neighbor_T=20}. For the range $k >1200$, all the plots with various $\gamma$ merge into a single line whose slope is close to 1. This shows that the simulated set is homogenous on large scales. In Fig.~\ref{fig: Dq_k-neighbor_T=20}, we computed the generalized dimension using several different scaling ranges of $k$. We can see that the result is sensitive to the choice of scale. The result from the near-neighbor method is included for comparison. Here, we used an identical range of $k$ for different $\gamma$. While the spectra computed with some ranges more nearly resemble the spectrum with the near-neighbor method, we were not able to find a range of $k$ where the generalized dimension closely traces the spectrum from the nearest neighbor method. For $q> -1$, the resulting values are compatible with those found using the BC and IC methods.

\begin{figure}[htbp]
\centering
\includegraphics[width=\textwidth]{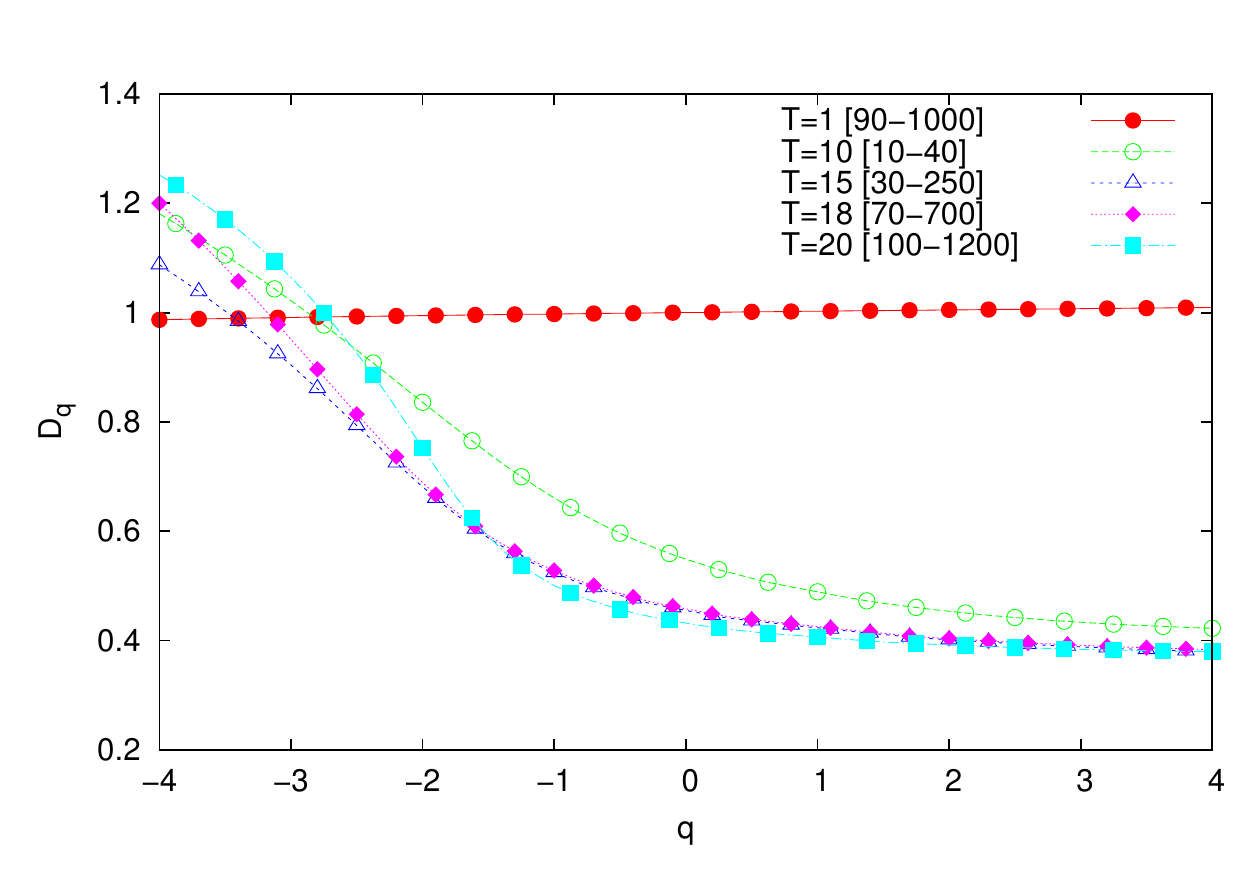}
\caption{{ This plot shows the generalized dimension $D_q$ for selected times $T$, computed with the $k$-neighbor method. }}
\label{fig: Dq_k-neighbor}
\end{figure}

\begin{figure}[htbp]
\centering
\includegraphics[width=\textwidth]{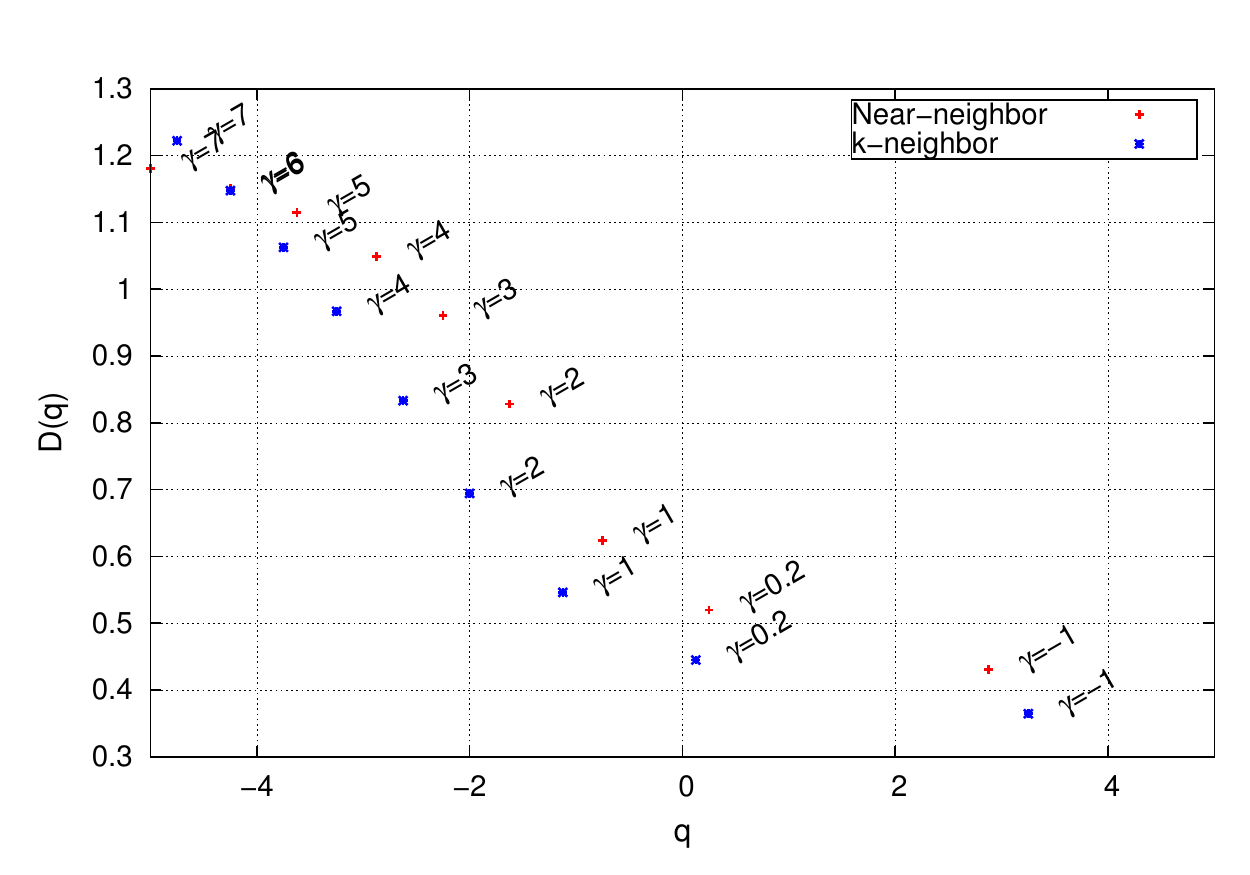}
\caption{{ This plot shows the values obtained for the generalized dimensions using different, mass-oriented, numerical methods. The corresponding value of $\gamma$ is also shown.}}
\label{fig: Dq_comparison}
\end{figure}

\begin{figure}[ht]
\centerline{\includegraphics[width=\textwidth]{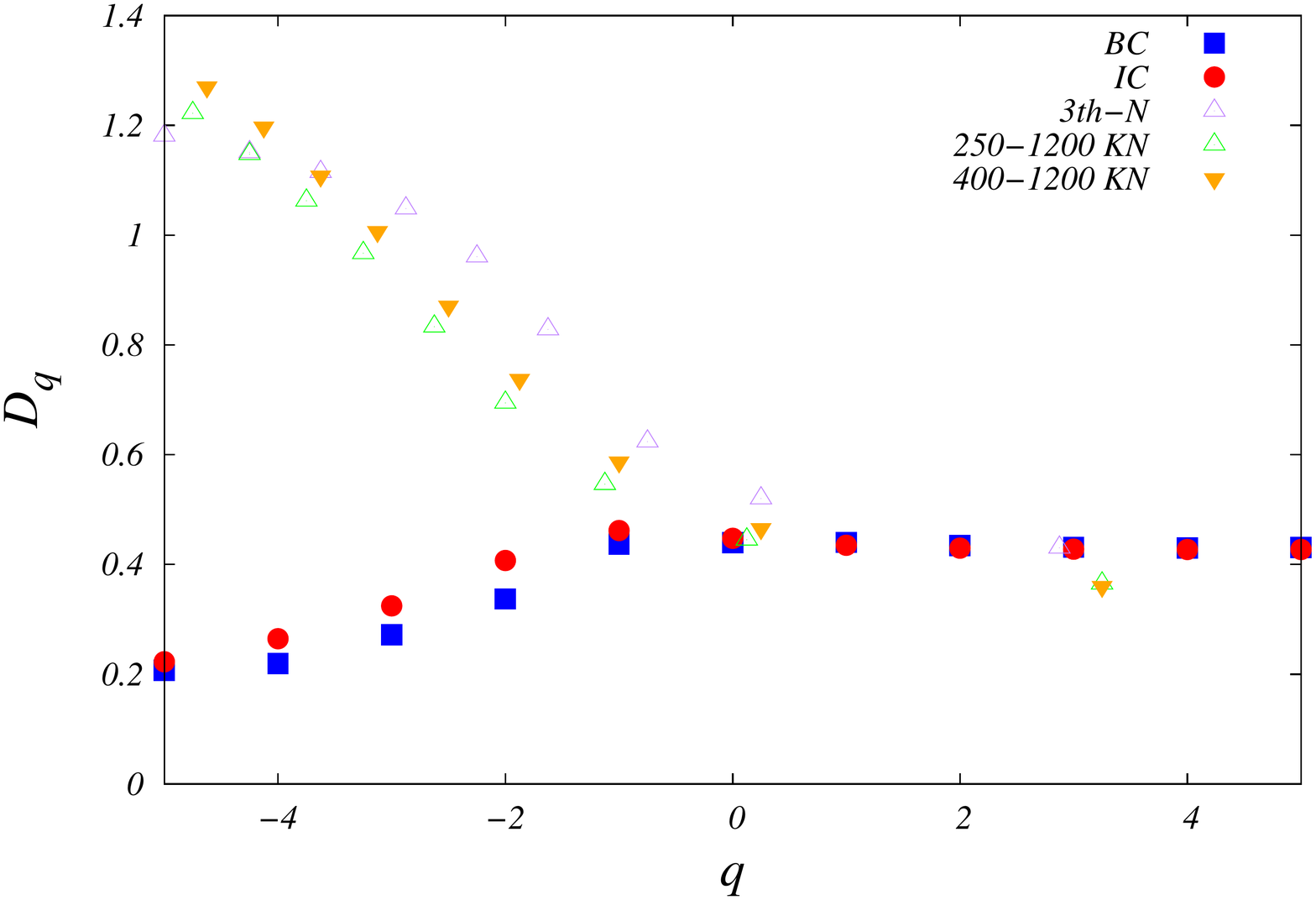}}
\caption{\label{fig6} Generalized dimensions $D_q(q)$ obtained at $T=20$  by the $k$-neighbor method (triangles). The results obtained by the BC and IC methods given in figure \ref{fig4} are shown for comparison.}
\end{figure}

\begin{figure}[ht]
\centerline{\includegraphics[width=\textwidth]{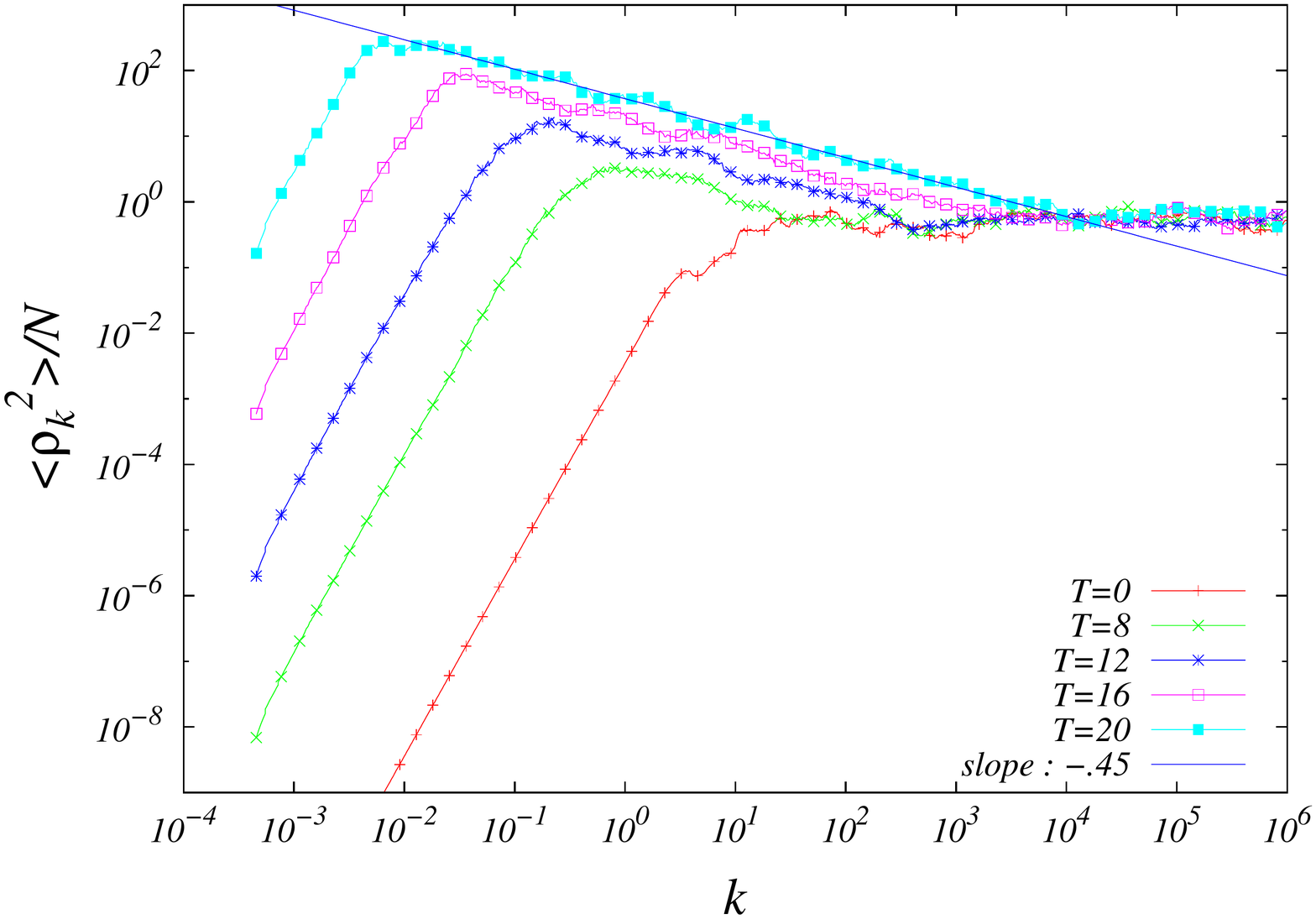}}
\caption{\label{fig7} Time evolution of the power spectrum of the density evaluated by averaging over 40 successive points. At $T=20$, in the interval $0,01<k<1000$, the slope $-0,45$ is obtained by linear regression.}
\end{figure}

It is useful to check values of $D_{q}$ by another method. For a fractal, the autocorrelation function of the density decays as a power law. From the autocorrelation function, or its Fourier transform which yields the power spectrum of the density fluctuations, it is possible to obtain the correlation dimension $D_2$.  We have the relation $D_2=-n$ where $n$ is the power of the wave number $k$ in the power spectrum of the density fluctuations \cite{millerrouet}. The power spectrum is shown in Figure \ref{fig7} at different times. At $T = 0 $, the initial condition imposes a slope of 3 for the longer wavelengths, which is verified in the figure. For large wave numbers ($k> k_c=2\pi/1$) the spectrum is 1, which reflects the lack of correlation between particles at separations of less than unity. Over time, one sees the formation of an intermediate scaling range that exhibits power law behavior. The size of the scaling region grows through decreasing its lower boundary and increasing the upper boundary: the slope is always the same. At $T=20$, a linear regression analysis in the range $.01<k<1000$ yields a slope $n=-0,45$ and therefore $D_2=0,45$. This value is quite consistent with those found by the IC,  BC and KN methods.

\section{Discussion and conclusion}\label{conclusion}

A simple one-dimensional model that only includes expansion and gravitation (Newtonian) and is subject to periodic boundary conditions was investigated numerically. This model belongs to a family of one-dimensional models of expansion  which only differ by their symmetry in the original space. The use of the quasi-invariance group \cite{Burgan}, for which the law of transformation of the spatial variable is different from that of the time unit, provides an autonomous equation of motion. Other choices are possible, such as taking the same transformation law ($A(t)=C(t)$). We then obtain a more standard form of the comoving coordinates. Friction disappears, but then a gravitational constant which depends explicitly on the time is introduced. \cite{peebles2}. This family of models, which describe the evolution of a planar disturbance, differ only in the value of the coefficient of friction. This coefficient approaches zero if we consider a high-dimensional $d$ space. Its properties were investigated elsewhere where it was referred to as the Hamiltonian case \cite{millerrouet}.

For three-dimensional simulations, it is necessary to introduce approximations in the gravitational field at both short and long distances. In particular, a short-range cut-off is necessary to maintain a manageable execution time. The influence of these approximations of the force law on the predicted fractal properties is not known, but could be significant. It is likely that they would limit the useful scaling range.  In one dimension, these approximations are not necessary. The equations of motion are exactly integrated between particle crossings so it is possible to write an event driven algorithm to follow the motion with the precision imposed by the word-length maintained by the computer for a set of N particles. This is important for fractal analysis that requires us to keep all possible precision for the particle trajectories. Also, in contrast with higher dimension, elsewhere we have shown that it is possible to analytically evaluate Ewald sums in 1D and thus exactly model periodic boundary conditions \cite{ewald}. As we have shown, these are the only boundary conditions that are consistent with a comoving frame.

The simulations were performed on an initially cold system with a density spectrum that follows a power law. This scale-free formulation is commonly used by astrophysicists \cite{Melott1} and is motivated by the generally accepted characteristics of the inflationary density field. Starting with a nearly homogeneous distribution of $N$ particles, simulations show a successive fractionation and recombination resulting in the formation of a hierarchical structure of clusters in both phase space and on the real line. After a sufficiently long time, only a single structure will remain. This is only due to the finite number of particles. However, even if the boundary effects become important by then, the structure of the virialized cluster thus formed could be explored, pointing to differences with virialization of an isolated particle system which arises only as a consequence of gravity (without expansion, and therefore without friction or a continuous background).

In earlier work we showed that the analysis by the BC and IC methods give a generalized dimension $D_q(q)$ which increases when q increases for $q<-1$, then remains almost constant for $q>1$ \cite{millerrouet}. However, a general property of the $D_q$ curve is that it should be non-increasing \cite{feder88}. Therefore, the results for $q<-1$ obtained by the BC and IC methods are questionable. As we have shown here, in contrast with the former methods based on partitions of equal size, the analysis based on partitions of equal mass or measure provides a generalized dimension $D_q(q)$ which decreases as $q$ in the low density regions. The fact that the results from the two distinct mass-oriented methods do not show complete agreement may be explained by the finite degree of hierarchy of the simulated set. Since the nearest neighbor method (NN) is concerned only with the local statistics of the given set, the method may extract  generalized dimensions associated with its subsets that can have different properties from its global properties. From the scaling plots of the $k$-neighbor method, we can see that scaling is observed in limited ranges which do not generally include small values, so $k=3$ is avoided. This may also contribute the the difference between the KN and near-neighbor methods since the latter explicitly employed $k=3$. For the high density regions, three methods, namely, the BC, IC and near-neighbor, are known to work well and they agree for $q>-1$.  In addition, the value of $D_2$  was extracted from the power spectrum, allowing for some confidence in the values obtained for $q>-1$.  Other work on known fractals indicates that mass-oriented approaches are superior for the analysis of low density regions, which dominate when $q$ is negative, or when $\gamma$ is positive \cite{yui}.

For observational data, Dubrulle and Lachieze-Rey attribute the growth of the $D_q(q)$ curve based on the BC or IC analysis to a lack of resolution. This argument can be considered here in two ways: Either the number of particles is not sufficient to properly represent low density areas or the analysis based on the configuration space alone is inadequate. As we have seen, the clusters actually form in phase, or $\mu$, space. The projected positions on the line cannot account for the existence of clusters of different speeds (in other words, one cluster can hide another). Work on the analysis of clusters in the phase space is currently in progress. For the RF model with a water bag type initial condition, previous work has shown that $D_0\sim.77$ \cite{Rouet2,MRGexp} . One approach for obtaining a better representation of low density regions is to use Eulerian simulations in which we follow the evolution of a distribution function in phase space. Work is also in progress with this approach.

In summary, we have seen that a one-dimensional model of an expanding universe  shares  qualitative properties with the observed universe \cite{Chacon}, such as the bottom-up scenario, where due to fragmentation, a set of structures come together to form larger structures and so on. This hierarchical structure has a multifractal character, a property found by the results of our simulations. In particular, for $q>1$ the generalized dimension is about 0.4 the function, while for $q<0$ the $D_q$  approaches about $1.4$ for $q$  about $-4$. This type of mutifractal behavior in a one-dimensional system is reminiscent of the well studied multiplicative binomial process \cite{feder88,yui}. The function $\tau_q$  exhibits two linear zones, one for $q>-1$, the other for $q<-1$. This suggests that the underlying geometry is bi-fractal, but caution must be excecersized because it reflects computations using partitions of equal size. Such a property was proposed some time ago by Balian and Schaeffer based on the analysis of data catalogs \cite{Bal}. The work also directly addresses the important question of the scale on which fractal behavior is observed at a given epoch. We have seen that the scaling range increases with time, thus limiting the size of the largest objects. Beyond a horizon that grows with time, the universe as modeled here, is truly homogenous. Thus, in contrast with some conjectures \cite{pietronero1987,Pietro_rev,Pietro2006,Pietro2002,cosfrac}, in the Einstein-de Sitter scenario the cosmological principle is verified, but the homogeneity scale cannot be viewed as constant and increasingly large structures, such as the Laniakea supercluster \cite{Tully:2014aa}, will develop as time progresses.

\acknowledgements

{The authors are grateful for the support of the Center for Information Services at Texas Christian University. They also benefitted from conversations with Serguei Tcheremchantsev and George Gilbert. B. Miller benefitted from support received from the Universit\'e d'Orl\'eans and J-L Rouet benefitted from support received from Texas Christian University.}

\bibliography{miller_rouet_shiozawa_v2}

\begin{thebibliography}{38}%
\makeatletter
\providecommand \@ifxundefined [1]{%
 \@ifx{#1\undefined}
}%
\providecommand \@ifnum [1]{%
 \ifnum #1\expandafter \@firstoftwo
 \else \expandafter \@secondoftwo
 \fi
}%
\providecommand \@ifx [1]{%
 \ifx #1\expandafter \@firstoftwo
 \else \expandafter \@secondoftwo
 \fi
}%
\providecommand \natexlab [1]{#1}%
\providecommand \enquote  [1]{``#1''}%
\providecommand \bibnamefont  [1]{#1}%
\providecommand \bibfnamefont [1]{#1}%
\providecommand \citenamefont [1]{#1}%
\providecommand \href@noop [0]{\@secondoftwo}%
\providecommand \href [0]{\begingroup \@sanitize@url \@href}%
\providecommand \@href[1]{\@@startlink{#1}\@@href}%
\providecommand \@@href[1]{\endgroup#1\@@endlink}%
\providecommand \@sanitize@url [0]{\catcode `\\12\catcode `\$12\catcode
  `\&12\catcode `\#12\catcode `\^12\catcode `\_12\catcode `\%12\relax}%
\providecommand \@@startlink[1]{}%
\providecommand \@@endlink[0]{}%
\providecommand \url  [0]{\begingroup\@sanitize@url \@url }%
\providecommand \@url [1]{\endgroup\@href {#1}{\urlprefix }}%
\providecommand \urlprefix  [0]{URL }%
\providecommand \Eprint [0]{\href }%
\providecommand \doibase [0]{http://dx.doi.org/}%
\providecommand \selectlanguage [0]{\@gobble}%
\providecommand \bibinfo  [0]{\@secondoftwo}%
\providecommand \bibfield  [0]{\@secondoftwo}%
\providecommand \translation [1]{[#1]}%
\providecommand \BibitemOpen [0]{}%
\providecommand \bibitemStop [0]{}%
\providecommand \bibitemNoStop [0]{.\EOS\space}%
\providecommand \EOS [0]{\spacefactor3000\relax}%
\providecommand \BibitemShut  [1]{\csname bibitem#1\endcsname}%
\let\auto@bib@innerbib\@empty
\bibitem [{\citenamefont {Peebles}(1993)}]{peebles2}%
  \BibitemOpen
  \bibfield  {author} {\bibinfo {author} {\bibfnamefont {P.~J.~E.}\
  \bibnamefont {Peebles}},\ }\href@noop {} {\emph {\bibinfo {title} {Principles
  of Physical Cosmology}}}\ (\bibinfo  {publisher} {Princeton University
  Press},\ \bibinfo {address} {Princeton, NJ},\ \bibinfo {year}
  {1993})\BibitemShut {NoStop}%
\bibitem [{\citenamefont {Tully}\ \emph {et~al.}(2014)\citenamefont {Tully},
  \citenamefont {Courtois}, \citenamefont {Hoffman},\ and\ \citenamefont
  {Pomarede}}]{Tully:2014aa}%
  \BibitemOpen
  \bibfield  {author} {\bibinfo {author} {\bibfnamefont {R.~B.}\ \bibnamefont
  {Tully}}, \bibinfo {author} {\bibfnamefont {H.}~\bibnamefont {Courtois}},
  \bibinfo {author} {\bibfnamefont {Y.}~\bibnamefont {Hoffman}}, \ and\
  \bibinfo {author} {\bibfnamefont {D.}~\bibnamefont {Pomarede}},\ }\href
  {http://dx.doi.org/10.1038/nature13674} {\bibfield  {journal} {\bibinfo
  {journal} {Nature}\ }\textbf {\bibinfo {volume} {513}},\ \bibinfo {pages}
  {71} (\bibinfo {year} {2014})}\BibitemShut {NoStop}%
\bibitem [{\citenamefont {Martinez}(2009)}]{Martinez}%
  \BibitemOpen
  \bibfield  {author} {\bibinfo {author} {\bibfnamefont {V.~J.}\ \bibnamefont
  {Martinez}},\ }\href@noop {} {\bibfield  {journal} {\bibinfo  {journal}
  {Lect. Notes Phys.}\ }\textbf {\bibinfo {volume} {665}},\ \bibinfo {pages}
  {269} (\bibinfo {year} {2009})}\BibitemShut {NoStop}%
\bibitem [{\citenamefont {Fang}(2006)}]{Fang}%
  \BibitemOpen
  \bibfield  {author} {\bibinfo {author} {\bibfnamefont {F.}~\bibnamefont
  {Fang}},\ }\href {http://stacks.iop.org/0004-637X/644/i=2/a=678} {\bibfield
  {journal} {\bibinfo  {journal} {The Astrophysical Journal}\ }\textbf
  {\bibinfo {volume} {644}},\ \bibinfo {pages} {678} (\bibinfo {year}
  {2006})}\BibitemShut {NoStop}%
\bibitem [{\citenamefont {Mandelbrot}(1982)}]{Man}%
  \BibitemOpen
  \bibfield  {author} {\bibinfo {author} {\bibfnamefont {B.}~\bibnamefont
  {Mandelbrot}},\ }\href@noop {} {\emph {\bibinfo {title} {The Fractal Geometry
  of Nature}}}\ (\bibinfo  {publisher} {Freeman},\ \bibinfo {address} {San
  Francisco},\ \bibinfo {year} {1982})\BibitemShut {NoStop}%
\bibitem [{\citenamefont {Feder}(1988)}]{feder88}%
  \BibitemOpen
  \bibfield  {author} {\bibinfo {author} {\bibfnamefont {J.}~\bibnamefont
  {Feder}},\ }\href@noop {} {\emph {\bibinfo {title} {Fractals}}}\ (\bibinfo
  {publisher} {Plenum Press},\ \bibinfo {address} {New York},\ \bibinfo {year}
  {1988})\BibitemShut {NoStop}%
\bibitem [{\citenamefont {Springiel}\ \emph {et~al.}(2006)\citenamefont
  {Springiel}, \citenamefont {Frenk},\ and\ \citenamefont {White}}]{Virgo}%
  \BibitemOpen
  \bibfield  {author} {\bibinfo {author} {\bibfnamefont {V.}~\bibnamefont
  {Springiel}}, \bibinfo {author} {\bibfnamefont {C.~S.}\ \bibnamefont
  {Frenk}}, \ and\ \bibinfo {author} {\bibfnamefont {S.~D.~M.}\ \bibnamefont
  {White}},\ }\href@noop {} {\bibfield  {journal} {\bibinfo  {journal}
  {Nature}\ }\textbf {\bibinfo {volume} {440}},\ \bibinfo {pages} {1137}
  (\bibinfo {year} {2006})}\BibitemShut {NoStop}%
\bibitem [{\citenamefont {Alimi}\ \emph {et~al.}(2012)\citenamefont {Alimi},
  \citenamefont {Bouillot}, \citenamefont {Rasera}, \citenamefont {Reverdy},
  \citenamefont {Corasaniti} \emph {et~al.}}]{Alimi:2012be}%
  \BibitemOpen
  \bibfield  {author} {\bibinfo {author} {\bibfnamefont {J.-M.}\ \bibnamefont
  {Alimi}}, \bibinfo {author} {\bibfnamefont {V.}~\bibnamefont {Bouillot}},
  \bibinfo {author} {\bibfnamefont {Y.}~\bibnamefont {Rasera}}, \bibinfo
  {author} {\bibfnamefont {V.}~\bibnamefont {Reverdy}}, \bibinfo {author}
  {\bibfnamefont {P.-S.}\ \bibnamefont {Corasaniti}},  \emph {et~al.},\
  }\href@noop {} {\bibfield  {journal} {\bibinfo  {journal} {arXiv}\ }
  (\bibinfo {year} {2012})},\ \Eprint {http://arxiv.org/abs/1206.2838}
  {arXiv:1206.2838 [astro-ph.CO]} \BibitemShut {NoStop}%
\bibitem [{\citenamefont {Rouet}\ \emph {et~al.}(1990)\citenamefont {Rouet},
  \citenamefont {Feix},\ and\ \citenamefont {Navet}}]{Rouet1}%
  \BibitemOpen
  \bibfield  {author} {\bibinfo {author} {\bibfnamefont {J.-L.}\ \bibnamefont
  {Rouet}}, \bibinfo {author} {\bibfnamefont {M.~R.}\ \bibnamefont {Feix}}, \
  and\ \bibinfo {author} {\bibfnamefont {M.}~\bibnamefont {Navet}},\ }in\
  \href@noop {} {\emph {\bibinfo {booktitle} {Vistas in Astronomy 33}}},\
  \bibinfo {editor} {edited by\ \bibinfo {editor} {\bibfnamefont
  {A.}~\bibnamefont {Heck}}}\ (\bibinfo  {publisher} {Pergamon},\ \bibinfo
  {year} {1990})\ pp.\ \bibinfo {pages} {357--370}\BibitemShut {NoStop}%
\bibitem [{\citenamefont {Fanelli}\ and\ \citenamefont
  {Aurell}(2002)}]{Fanelli}%
  \BibitemOpen
  \bibfield  {author} {\bibinfo {author} {\bibfnamefont {D.}~\bibnamefont
  {Fanelli}}\ and\ \bibinfo {author} {\bibfnamefont {E.}~\bibnamefont
  {Aurell}},\ }\href@noop {} {\bibfield  {journal} {\bibinfo  {journal}
  {Astronomy and Astrophysics}\ }\textbf {\bibinfo {volume} {395}},\ \bibinfo
  {pages} {399} (\bibinfo {year} {2002})}\BibitemShut {NoStop}%
\bibitem [{\citenamefont {Gabrielli}\ \emph {et~al.}(2009)\citenamefont
  {Gabrielli}, \citenamefont {Joyce},\ and\ \citenamefont {Sicard}}]{Joyce_1d}%
  \BibitemOpen
  \bibfield  {author} {\bibinfo {author} {\bibfnamefont {A.}~\bibnamefont
  {Gabrielli}}, \bibinfo {author} {\bibfnamefont {M.}~\bibnamefont {Joyce}}, \
  and\ \bibinfo {author} {\bibfnamefont {F.}~\bibnamefont {Sicard}},\
  }\href@noop {} {\bibfield  {journal} {\bibinfo  {journal} {Physical Review
  E}\ }\textbf {\bibinfo {volume} {80}},\ \bibinfo {pages} {041108} (\bibinfo
  {year} {2009})}\BibitemShut {NoStop}%
\bibitem [{\citenamefont {Miller}\ and\ \citenamefont
  {Rouet}(2010{\natexlab{a}})}]{millerrouet}%
  \BibitemOpen
  \bibfield  {author} {\bibinfo {author} {\bibfnamefont {B.~N.}\ \bibnamefont
  {Miller}}\ and\ \bibinfo {author} {\bibfnamefont {J.-L.}\ \bibnamefont
  {Rouet}},\ }\href {\doibase 10.1088/1742-5468/2010/12/P12028} {\bibfield
  {journal} {\bibinfo  {journal} {JSTAT}\ }\textbf {\bibinfo {volume} {2010}}
  (\bibinfo {year} {2010}{\natexlab{a}}),\
  10.1088/1742-5468/2010/12/P12028}\BibitemShut {NoStop}%
\bibitem [{\citenamefont {Miller}\ and\ \citenamefont
  {Rouet}(2010{\natexlab{b}})}]{ewald}%
  \BibitemOpen
  \bibfield  {author} {\bibinfo {author} {\bibfnamefont {B.~N.}\ \bibnamefont
  {Miller}}\ and\ \bibinfo {author} {\bibfnamefont {J.-L.}\ \bibnamefont
  {Rouet}},\ }\href {\doibase 10.1103/PhysRevE.82.066203} {\bibfield  {journal}
  {\bibinfo  {journal} {Phys. Rev. E}\ }\textbf {\bibinfo {volume} {82}},\
  \bibinfo {pages} {066203} (\bibinfo {year} {2010}{\natexlab{b}})}\BibitemShut
  {NoStop}%
\bibitem [{\citenamefont {{Hohl}}\ and\ \citenamefont
  {{Feix}}(1967)}]{HohlFeix}%
  \BibitemOpen
  \bibfield  {author} {\bibinfo {author} {\bibfnamefont {F.}~\bibnamefont
  {{Hohl}}}\ and\ \bibinfo {author} {\bibfnamefont {M.~R.}\ \bibnamefont
  {{Feix}}},\ }\href {\doibase 10.1086/149106} {\bibfield  {journal} {\bibinfo
  {journal} {The Astrophysical Journal}\ }\textbf {\bibinfo {volume} {147}},\
  \bibinfo {pages} {1164} (\bibinfo {year} {1967})}\BibitemShut {NoStop}%
\bibitem [{\citenamefont {Yawn}\ and\ \citenamefont {Miller}(2003)}]{Yawn}%
  \BibitemOpen
  \bibfield  {author} {\bibinfo {author} {\bibfnamefont {K.~R.}\ \bibnamefont
  {Yawn}}\ and\ \bibinfo {author} {\bibfnamefont {B.~N.}\ \bibnamefont
  {Miller}},\ }\href {\doibase 10.1103/PhysRevE.68.056120} {\bibfield
  {journal} {\bibinfo  {journal} {Physical Review E}\ }\textbf {\bibinfo
  {volume} {68}},\ \bibinfo {pages} {056120} (\bibinfo {year}
  {2003})}\BibitemShut {NoStop}%
\bibitem [{\citenamefont {Tsuchiya}\ \emph {et~al.}(1996)\citenamefont
  {Tsuchiya}, \citenamefont {Konishi},\ and\ \citenamefont
  {Gouda}}]{tsuchiya2}%
  \BibitemOpen
  \bibfield  {author} {\bibinfo {author} {\bibfnamefont {T.}~\bibnamefont
  {Tsuchiya}}, \bibinfo {author} {\bibfnamefont {T.}~\bibnamefont {Konishi}}, \
  and\ \bibinfo {author} {\bibfnamefont {N.}~\bibnamefont {Gouda}},\
  }\href@noop {} {\bibfield  {journal} {\bibinfo  {journal} {Phys. Rev. E}\
  }\textbf {\bibinfo {volume} {53}},\ \bibinfo {pages} {2210} (\bibinfo {year}
  {1996})}\BibitemShut {NoStop}%
\bibitem [{\citenamefont {Miller}\ \emph {et~al.}(2007)\citenamefont {Miller},
  \citenamefont {Rouet},\ and\ \citenamefont {Guirriec}}]{MRGexp}%
  \BibitemOpen
  \bibfield  {author} {\bibinfo {author} {\bibfnamefont {B.~N.}\ \bibnamefont
  {Miller}}, \bibinfo {author} {\bibfnamefont {J.-L.}\ \bibnamefont {Rouet}}, \
  and\ \bibinfo {author} {\bibfnamefont {E.~L.}\ \bibnamefont {Guirriec}},\
  }\href@noop {} {\bibfield  {journal} {\bibinfo  {journal} {Physical Review
  E}\ }\textbf {\bibinfo {volume} {76}},\ \bibinfo {pages} {036705} (\bibinfo
  {year} {2007})}\BibitemShut {NoStop}%
\bibitem [{\citenamefont {Besnard}\ \emph {et~al.}(1983)\citenamefont
  {Besnard}, \citenamefont {Burgan}, \citenamefont {Munier}, \citenamefont
  {Feix},\ and\ \citenamefont {Fijalkow}}]{besnard:1123}%
  \BibitemOpen
  \bibfield  {author} {\bibinfo {author} {\bibfnamefont {D.}~\bibnamefont
  {Besnard}}, \bibinfo {author} {\bibfnamefont {J.~R.}\ \bibnamefont {Burgan}},
  \bibinfo {author} {\bibfnamefont {A.}~\bibnamefont {Munier}}, \bibinfo
  {author} {\bibfnamefont {M.~R.}\ \bibnamefont {Feix}}, \ and\ \bibinfo
  {author} {\bibfnamefont {E.}~\bibnamefont {Fijalkow}},\ }\href {\doibase
  10.1063/1.525839} {\bibfield  {journal} {\bibinfo  {journal} {Journal of
  Mathematical Physics}\ }\textbf {\bibinfo {volume} {24}},\ \bibinfo {pages}
  {1123} (\bibinfo {year} {1983})}\BibitemShut {NoStop}%
\bibitem [{\citenamefont {chan Hwang}\ and\ \citenamefont
  {Noh}(2006)}]{Newtap}%
  \BibitemOpen
  \bibfield  {author} {\bibinfo {author} {\bibfnamefont {J.}~\bibnamefont {chan
  Hwang}}\ and\ \bibinfo {author} {\bibfnamefont {H.}~\bibnamefont {Noh}},\
  }\href@noop {} {\bibfield  {journal} {\bibinfo  {journal} {MNRAS}\ ,\
  \bibinfo {pages} {in press}} (\bibinfo {year} {2006})}\BibitemShut {NoStop}%
\bibitem [{\citenamefont {Bertschinger}(1998)}]{Ber}%
  \BibitemOpen
  \bibfield  {author} {\bibinfo {author} {\bibfnamefont {E.}~\bibnamefont
  {Bertschinger}},\ }\href@noop {} {\bibfield  {journal} {\bibinfo  {journal}
  {Annual Review of Astronomy and Astrophysics}\ }\textbf {\bibinfo {volume}
  {36}},\ \bibinfo {pages} {599} (\bibinfo {year} {1998})}\BibitemShut
  {NoStop}%
\bibitem [{\citenamefont {Peacock}(1999)}]{peacock}%
  \BibitemOpen
  \bibfield  {author} {\bibinfo {author} {\bibfnamefont {J.~A.}\ \bibnamefont
  {Peacock}},\ }\href@noop {} {\emph {\bibinfo {title} {Cosmological
  Physics}}}\ (\bibinfo  {publisher} {Cambridge},\ \bibinfo {address}
  {Cambridge},\ \bibinfo {year} {1999})\BibitemShut {NoStop}%
\bibitem [{\citenamefont {Badii}\ and\ \citenamefont {Politi}(1985)}]{Badii85}%
  \BibitemOpen
  \bibfield  {author} {\bibinfo {author} {\bibfnamefont {R.}~\bibnamefont
  {Badii}}\ and\ \bibinfo {author} {\bibfnamefont {A.}~\bibnamefont {Politi}},\
  }\href {\doibase 10.1007/BF01009897} {\bibfield  {journal} {\bibinfo
  {journal} {Journal of Statistical Physics}\ }\textbf {\bibinfo {volume}
  {40}},\ \bibinfo {pages} {725} (\bibinfo {year} {1985})}\BibitemShut
  {NoStop}%
\bibitem [{\citenamefont {van~de Water}\ and\ \citenamefont
  {Schram}(1988)}]{Water88}%
  \BibitemOpen
  \bibfield  {author} {\bibinfo {author} {\bibfnamefont {W.}~\bibnamefont
  {van~de Water}}\ and\ \bibinfo {author} {\bibfnamefont {P.}~\bibnamefont
  {Schram}},\ }\href {\doibase 10.1103/PhysRevA.37.3118} {\bibfield  {journal}
  {\bibinfo  {journal} {Phys. Rev. A}\ }\textbf {\bibinfo {volume} {37}},\
  \bibinfo {pages} {3118} (\bibinfo {year} {1988})}\BibitemShut {NoStop}%
\bibitem [{\citenamefont {Shiozawa}\ \emph {et~al.}(2014)\citenamefont
  {Shiozawa}, \citenamefont {Miller},\ and\ \citenamefont {Rouet}}]{yui}%
  \BibitemOpen
  \bibfield  {author} {\bibinfo {author} {\bibfnamefont {Y.}~\bibnamefont
  {Shiozawa}}, \bibinfo {author} {\bibfnamefont {B.~N.}\ \bibnamefont
  {Miller}}, \ and\ \bibinfo {author} {\bibfnamefont {J.-L.}\ \bibnamefont
  {Rouet}},\ }\href@noop {} {\bibfield  {journal} {\bibinfo  {journal} {CHAOS}\
  }\textbf {\bibinfo {volume} {24}},\ \bibinfo {pages} {033106} (\bibinfo
  {year} {2014})}\BibitemShut {NoStop}%
\bibitem [{\citenamefont {Riedi}(1995)}]{Riedi95}%
  \BibitemOpen
  \bibfield  {author} {\bibinfo {author} {\bibfnamefont {R.}~\bibnamefont
  {Riedi}},\ }\href {\doibase http://dx.doi.org/10.1006/jmaa.1995.1030}
  {\bibfield  {journal} {\bibinfo  {journal} {Journal of Mathematical Analysis
  and Applications}\ }\textbf {\bibinfo {volume} {189}},\ \bibinfo {pages} {462
  } (\bibinfo {year} {1995})}\BibitemShut {NoStop}%
\bibitem [{\citenamefont {Grassberger}\ and\ \citenamefont
  {Procaccia}(1983)}]{Itamar83}%
  \BibitemOpen
  \bibfield  {author} {\bibinfo {author} {\bibfnamefont {P.}~\bibnamefont
  {Grassberger}}\ and\ \bibinfo {author} {\bibfnamefont {I.}~\bibnamefont
  {Procaccia}},\ }\href@noop {} {\bibfield  {journal} {\bibinfo  {journal}
  {Physical review letters}\ }\textbf {\bibinfo {volume} {50}},\ \bibinfo
  {pages} {346} (\bibinfo {year} {1983})}\BibitemShut {NoStop}%
\bibitem [{\citenamefont {Abramowitz}(1970)}]{abramowitz70}%
  \BibitemOpen
  \bibfield  {author} {\bibinfo {author} {\bibfnamefont {M.}~\bibnamefont
  {Abramowitz}},\ }\href@noop {} {\emph {\bibinfo {title} {Handbook of
  mathematical functions}}}\ (\bibinfo  {publisher} {Dover Publications},\
  \bibinfo {address} {New York},\ \bibinfo {year} {1970})\BibitemShut {NoStop}%
\bibitem [{\citenamefont {Broggi}(1988)}]{broggi1988evaluation}%
  \BibitemOpen
  \bibfield  {author} {\bibinfo {author} {\bibfnamefont {G.}~\bibnamefont
  {Broggi}},\ }\href@noop {} {\bibfield  {journal} {\bibinfo  {journal} {JOSA
  B}\ }\textbf {\bibinfo {volume} {5}},\ \bibinfo {pages} {1020} (\bibinfo
  {year} {1988})}\BibitemShut {NoStop}%
\bibitem [{\citenamefont {Burgan}\ \emph {et~al.}(1978)\citenamefont {Burgan},
  \citenamefont {Gutierrez}, \citenamefont {Munier}, \citenamefont {Fijalkow},\
  and\ \citenamefont {Feix}}]{Burgan}%
  \BibitemOpen
  \bibfield  {author} {\bibinfo {author} {\bibfnamefont {J.}~\bibnamefont
  {Burgan}}, \bibinfo {author} {\bibfnamefont {J.}~\bibnamefont {Gutierrez}},
  \bibinfo {author} {\bibfnamefont {A.}~\bibnamefont {Munier}}, \bibinfo
  {author} {\bibfnamefont {E.}~\bibnamefont {Fijalkow}}, \ and\ \bibinfo
  {author} {\bibfnamefont {M.}~\bibnamefont {Feix}},\ }in\ \href@noop {} {\emph
  {\bibinfo {booktitle} {Strongly Coupled Plasmas}}},\ \bibinfo {editor}
  {edited by\ \bibinfo {editor} {\bibfnamefont {K.}~\bibnamefont {Gabor}}\ and\
  \bibinfo {editor} {\bibfnamefont {C.}~\bibnamefont {Paul}}}\ (\bibinfo
  {publisher} {Plenum Press},\ \bibinfo {address} {New York, USA},\ \bibinfo
  {year} {1978})\ pp.\ \bibinfo {pages} {597--641}\BibitemShut {NoStop}%
\bibitem [{\citenamefont {Adrian~L.}\ \emph {et~al.}(1983)\citenamefont
  {Adrian~L.}, \citenamefont {Jaan}, \citenamefont {Ivar}, \citenamefont
  {Anatoli~A.},\ and\ \citenamefont {Shandarin}}]{Melott1}%
  \BibitemOpen
  \bibfield  {author} {\bibinfo {author} {\bibfnamefont {M.}~\bibnamefont
  {Adrian~L.}}, \bibinfo {author} {\bibfnamefont {E.}~\bibnamefont {Jaan}},
  \bibinfo {author} {\bibfnamefont {S.}~\bibnamefont {Ivar}}, \bibinfo {author}
  {\bibfnamefont {K.}~\bibnamefont {Anatoli~A.}}, \ and\ \bibinfo {author}
  {\bibfnamefont {S.~F.}\ \bibnamefont {Shandarin}},\ }\href@noop {} {\bibfield
   {journal} {\bibinfo  {journal} {Physical Review Letters}\ }\textbf {\bibinfo
  {volume} {51}},\ \bibinfo {pages} {935} (\bibinfo {year} {1983})}\BibitemShut
  {NoStop}%
\bibitem [{\citenamefont {Rouet}\ \emph {et~al.}(1991)\citenamefont {Rouet},
  \citenamefont {Jamin},\ and\ \citenamefont {Feix}}]{Rouet2}%
  \BibitemOpen
  \bibfield  {author} {\bibinfo {author} {\bibfnamefont {J.-L.}\ \bibnamefont
  {Rouet}}, \bibinfo {author} {\bibfnamefont {E.}~\bibnamefont {Jamin}}, \ and\
  \bibinfo {author} {\bibfnamefont {M.~R.}\ \bibnamefont {Feix}},\ }in\
  \href@noop {} {\emph {\bibinfo {booktitle} {Applying Fractals in
  Astronomy}}},\ \bibinfo {editor} {edited by\ \bibinfo {editor} {\bibfnamefont
  {A.}~\bibnamefont {Heck}}\ and\ \bibinfo {editor} {\bibfnamefont {J.~M.}\
  \bibnamefont {Perdang}}}\ (\bibinfo  {publisher} {Springer-Verlag, Berlin},\
  \bibinfo {year} {1991})\ pp.\ \bibinfo {pages} {161--179}\BibitemShut
  {NoStop}%
\bibitem [{\citenamefont {Chacon-Cardona}\ and\ \citenamefont
  {Casas-Miranda}(2012)}]{Chacon}%
  \BibitemOpen
  \bibfield  {author} {\bibinfo {author} {\bibfnamefont {C.~A.}\ \bibnamefont
  {Chacon-Cardona}}\ and\ \bibinfo {author} {\bibfnamefont {R.~A.}\
  \bibnamefont {Casas-Miranda}},\ }\href@noop {} {\bibfield  {journal}
  {\bibinfo  {journal} {arXiv}\ } (\bibinfo {year} {2012})},\ \Eprint
  {http://arxiv.org/abs/1212.4832} {arXiv:1212.4832 [astro-ph.CO]} \BibitemShut
  {NoStop}%
\bibitem [{\citenamefont {{Balian}}\ and\ \citenamefont
  {{Schaeffer}}(1989)}]{Bal}%
  \BibitemOpen
  \bibfield  {author} {\bibinfo {author} {\bibfnamefont {R.}~\bibnamefont
  {{Balian}}}\ and\ \bibinfo {author} {\bibfnamefont {R.}~\bibnamefont
  {{Schaeffer}}},\ }\href@noop {} {\bibfield  {journal} {\bibinfo  {journal}
  {Astronomy and Astrophysics}\ }\textbf {\bibinfo {volume} {226}},\ \bibinfo
  {pages} {373} (\bibinfo {year} {1989})}\BibitemShut {NoStop}%
\bibitem [{\citenamefont {Pietronero}(1987)}]{pietronero1987}%
  \BibitemOpen
  \bibfield  {author} {\bibinfo {author} {\bibfnamefont {L.}~\bibnamefont
  {Pietronero}},\ }\href@noop {} {\bibfield  {journal} {\bibinfo  {journal}
  {Physica A}\ }\textbf {\bibinfo {volume} {144}},\ \bibinfo {pages} {257}
  (\bibinfo {year} {1987})}\BibitemShut {NoStop}%
\bibitem [{\citenamefont {Labini}\ \emph {et~al.}(1998)\citenamefont {Labini},
  \citenamefont {Montuori},\ and\ \citenamefont {Pietronero}}]{Pietro_rev}%
  \BibitemOpen
  \bibfield  {author} {\bibinfo {author} {\bibfnamefont {F.~S.}\ \bibnamefont
  {Labini}}, \bibinfo {author} {\bibfnamefont {M.}~\bibnamefont {Montuori}}, \
  and\ \bibinfo {author} {\bibfnamefont {L.}~\bibnamefont {Pietronero}},\
  }\href@noop {} {\bibfield  {journal} {\bibinfo  {journal} {Physics Reports}\
  }\textbf {\bibinfo {volume} {293}},\ \bibinfo {pages} {61} (\bibinfo {year}
  {1998})}\BibitemShut {NoStop}%
\bibitem [{\citenamefont {Pietronero}\ and\ \citenamefont
  {Labini}(2006)}]{Pietro2006}%
  \BibitemOpen
  \bibfield  {author} {\bibinfo {author} {\bibfnamefont {L.}~\bibnamefont
  {Pietronero}}\ and\ \bibinfo {author} {\bibfnamefont {F.~S.}\ \bibnamefont
  {Labini}},\ }\href@noop {} {\bibfield  {journal} {\bibinfo  {journal}
  {European Physics Journal B}\ }\textbf {\bibinfo {volume} {50}},\ \bibinfo
  {pages} {285} (\bibinfo {year} {2006})}\BibitemShut {NoStop}%
\bibitem [{\citenamefont {Montuori}\ and\ \citenamefont
  {Pietronero}(2002)}]{Pietro2002}%
  \BibitemOpen
  \bibfield  {author} {\bibinfo {author} {\bibfnamefont {M.}~\bibnamefont
  {Montuori}}\ and\ \bibinfo {author} {\bibfnamefont {L.}~\bibnamefont
  {Pietronero}},\ }in\ \href@noop {} {\emph {\bibinfo {booktitle} {Modern
  cosmology}}},\ \bibinfo {editor} {edited by\ \bibinfo {editor} {\bibfnamefont
  {S.}~\bibnamefont {Bonometto}}, \bibinfo {editor} {\bibfnamefont
  {V.}~\bibnamefont {Gorini}}, \ and\ \bibinfo {editor} {\bibfnamefont
  {U.}~\bibnamefont {Moschella}}}\ (\bibinfo  {publisher} {Institute of Physics
  Publishing},\ \bibinfo {address} {Bristol, UK},\ \bibinfo {year} {2002})\
  pp.\ \bibinfo {pages} {367--377}\BibitemShut {NoStop}%
\bibitem [{\citenamefont {Baryshev}\ and\ \citenamefont
  {Teerikorpi}(2002)}]{cosfrac}%
  \BibitemOpen
  \bibfield  {author} {\bibinfo {author} {\bibfnamefont {Y.}~\bibnamefont
  {Baryshev}}\ and\ \bibinfo {author} {\bibfnamefont {P.}~\bibnamefont
  {Teerikorpi}},\ }\href@noop {} {\emph {\bibinfo {title} {Discovery of Cosmic
  Fractals}}}\ (\bibinfo  {publisher} {World Scientific},\ \bibinfo {address}
  {Singapore},\ \bibinfo {year} {2002})\BibitemShut {NoStop}%
\end{thebibliography}%

\end{document}